\documentclass[12pt]{article}
\usepackage{amsthm, amsmath, amssymb}
\usepackage{latexsym,amsmath,amsfonts,amsthm,amssymb}
\input amssym.def
\input amssym.tex

\newtheorem{theorem}{Theorem}[section]

\usepackage{bbm}
\usepackage{tikz}
\usepackage[dvips]{epsfig}
\usepackage{graphicx}

\usepackage[font={footnotesize},margin=1cm]{caption}

\usepackage[breaklinks=false,pdfpagemode=None,pdfview=FitH,pdfstartview=FitH,citebordercolor={0 0 1},linkbordercolor={0 0 1},urlbordercolor={0 0 1},pagebordercolor={0 0 1},pdfborder={0 0 1}]{hyperref}

\usetikzlibrary{arrows,shapes}
\usetikzlibrary{decorations.pathmorphing}

\newcommand{\calD}{{\mathcal{D}}}
\newcommand{\calE}{{\mathcal{E}}}
\newcommand{\calF}{{\mathcal{F}}}

\newcommand{\per}{\mathrm{per}}
\newcommand{\res}{\mathrm{res}}
\newcommand{\ext}{\mathrm{ext}}
\newcommand{\prov}{\mathrm{prov}}
\newcommand{\tr}{\mathrm{tr}}
\newcommand{\sint}{\mathrm{sint}}
\newcommand{\lint}{\mathrm{lint}}
\newcommand{\ad}{\mathrm{ad}\,}
\newcommand{\ddd}{\mathrm{d}}
\newcommand{\sss}{\mathrm{s}}
\newcommand{\intint}{\mathrm{int}}

\numberwithin{equation}{section}

\newtheorem{proposition}[theorem]{Proposition}
\title  {
        Multi-Scale Jacobi Method for Anderson Localization
                 }

\author{
John Z.\ Imbrie\footnote{
This research was conducted in part while
the author was visiting the Institute for Advanced Study in Princeton, supported by The Fund for Math and The Ellentuck Fund.
}
\\Department of Mathematics,
University of Virginia \\
Charlottesville, VA 22904-4137, USA
\\ {\tt imbrie@virginia.edu}
}
\date{}
\begin{document}
\maketitle

\begin{abstract}
A new KAM-style proof of Anderson localization is obtained. A sequence of local rotations
is defined, such that off-diagonal matrix elements of the Hamiltonian are driven rapidly to zero. This leads to the first proof via multi-scale analysis of exponential decay of the eigenfunction correlator (this implies strong dynamical localization). The method has been used in recent work on many-body localization \cite{Imbrie2014}.
\end{abstract}
%\tableofcontents
\section{Introduction}\label{1}
%\subsection{Background}\label{1.1}
\noindent 
This work presents a new proof of localization in the Anderson tight-binding model at large disorder. In the spirit of KAM, a sequence of local rotations is used to diagonalize the Hamiltonian. This contrasts with previous work, which has largely focused on proving properties of the resolvent. Here we work directly with the eigenfunctions. We prove exponential decay of the eigenfunction correlator 
$\mathbb{E}\sum_\alpha\big|\psi_\alpha(x)\psi_\alpha(y)\big|$. 
Then strong dynamical localization is an immediate consequence. This work was motivated by a desire to find a procedure that might generalize to many-body Hamiltonians. We have successfully applied these ideas to a proof of many-body localization for a one-dimensional spin chain, under a certain assumption on level statistics \cite{Imbrie2014}. The key to success in the many-body context is exponential bounds on probabilities, for example the probability that 
$\sum_\alpha\big|\psi_\alpha(x)\psi_\alpha(y)\big|$
 is \textit{not} exponentially small. Such bounds have been proven by working with fractional moments of the resolvent \cite{Aizenman1994}, but here we present the first proof using multi-scale analysis. We have avoided using resolvent methods in this work because they do not seem to generalize to many-body problems.

Consider the random Schr\"odinger operator on $\mathbb{Z}^{D}$:
\begin{equation}
H=J_0(-\Delta-2D)+V. \label{(1.1)}
\end{equation}
Working on a rectangular subset $\Lambda \subset \mathbb Z^{D}$, the Hamiltonian is an operator on $\mathbb{C}^{|\Lambda|}$:
\begin{equation}
H_{xy}^{(\Lambda)}=\begin{cases}
v_x, &x=y; \\
-J_0, &|x-y|=1; \\
0, &\text{otherwise.}
\end{cases}\label{(1.2)}
\end{equation}
Here we use the $L^1$ distance $|x|=\sum^{D}_{\mu=1} |x_{\mu}|$ for $x=(x_1, \ldots ,
x_{D})\in \mathbb{Z}^{D}$.
The potentials $v_x$ are iid random variables with a fixed, continuous distribution having
a bounded density with respect to Lebesgue measure:
\begin{equation}
d \lambda(v)=\rho(v)dv, \label{(1.3)}
\end{equation}
with
\begin{equation}
\sup \limits_{v \in \mathbb R} \rho(v) \equiv \rho_0. \label{(1.4)}
\end{equation}
We prove exponential decay of the eigenfunction correlator for small $J_0$, with
bounds uniform in $\Lambda$.
\begin{theorem} \label{thm:1.1}
The eigenvalues of $H^{(\Lambda)}$ are nondegenerate, with probability 1. Let
$\{ \psi_\alpha(x)\}_{\alpha = 1,\ldots,|\Lambda|}$ denote the associated eigenvectors.
There is a $\kappa>0$ such that if $J_0$ is sufficiently small (depending only
on $D$ and $\rho_0$),  the following bounds hold for any rectangle $\Lambda$. The eigenfunction correlator satisfies
\begin{equation}
\mathbb{E} \, \sum\limits_{\alpha} \big|\psi_{\alpha}(x)\psi_{\alpha}(y)\big|\ \leq J_0^{\kappa|x-y|},
\label{(1.5)}
\end{equation}
and consequently, 
\begin{equation}
\sum\limits_{\alpha}\big|\psi_{\alpha}(x)\psi_{\alpha}(y)\big| \leq
J_0^{\kappa|x-y|/2} \ \text{with probability} \ 1-J_0^{\kappa|x-y|/2}. \label{(1.6)}  
\end{equation}
\end{theorem}

Dynamical localization refers to the rapid fall-off of $\sup_t \big|(e^{itH}P_I)(x,y)\big|$ with $|x-y|$, where $P_I$
is the projection onto some energy interval $I$. In the strong form, one has rapid decay of 
$\mathbb{E}\sup_t\big|(e^{itH}P_I)(x,y)\big|$. Previous work has followed one of two paths. The multi-scale analysis program began with proof of absence of diffusion via analysis of resonant regions and associated bounds on the resolvent \cite{Frohlich1983}. Subsequent work established dynamical localization in various forms by relating properties of the resolvent to properties of the 
eigenfunctions \cite{Martinelli1987,Germinet1998,Damanik2001}. 
The best result was of the form
$\mathbb{E}\sup_t\big|(e^{itH}P_I)(x,y)\big| \le \exp(-|x-y|^\zeta)$ for $\zeta < 1$ 
\cite{Germinet2001}. The dominant contribution to these bounds comes from 
probabilities of resonant regions. The fractional moment method began with a proof of exponential decay of $\mathbb{E}|(H-E)^{-1}(x,y)|^s$ for some $s<1$ 
\cite{Aizenman1993}. Subsequent work used this result to obtain exponential decay of
$\mathbb{E}\sup_t\big|(e^{itH}P_I)(x,y)\big|$, thereby obtaining dynamical localization in the 
strongest form \cite{Aizenman1994,Aizenman1998,Hundertmark2000,Aizenman2001}.

Implicit in these results are bounds proving the rapid fall-off of the eigenfunction correlator
$\mathbb{E}\sum_{E_\alpha\in I}\big|\psi_\alpha(x)\psi_\alpha(y)\big|$, from which one obtains dynamical localization by bounding $e^{itE_\alpha}$ by 1. Chulaevsky has 
developed a hybrid approach \cite{Chulaevsky2012,Chulaevsky2014a} with a greater
focus on eigenfunction correlators.

In this work we focus on the unitary rotations that diagonalize the Hamiltonian.
The columns of these rotations are the eigenfunctions. The rotation matrices are never singular, unlike the resolvent, which has poles at the eigenvalues. As a result, we are able to work with very mild separation conditions between resonant regions. This makes it possible to preserve exponential decay of probabilities of resonant regions. Exponential decay of probabilities is a critical requirement for progressing to many-body Hamiltonians, because the number of transitions possible in a region of size $n$ is exponential in $n$. 

We work on a sequence of length scales $L_k = (\frac{15}{8})^k$, designing rotations that connect sites separated by distances of the order of $L_k$. In nonresonant regions, the rotations are written as convergent power series based on first-order perturbation theory.
In resonant regions (blocks for quasi-degenerate perturbation theory), exact rotations are used, as in Jacobi diagonalization 
\cite{Sleijpen2000}. The procedure leads to rapid convergence to a diagonal
Hamiltonian, with off-diagonal matrix elements $\lesssim J_0^{L_k}$. As the unperturbed
eigenstates are deformed into the exact ones, we obtain a one-to-one mapping of
eigenstates to sites (except in rare resonant regions, where $n$ states map to $n$ sites).
The end result is a set of convergent graphical expansions for the eigenvalues and eigenfunctions, with each graph depending on the potential only in a neighborhood of its support. The detailed, local control of eigenvalues and eigenfunctions allows us to prove convergence in the $\Lambda \rightarrow \infty$ limit. The expansions should be useful for a more detailed analysis of their behavior in both energy space and position space.

Previous authors have used KAM-type procedures in the context of quasiperiodic and deterministic potentials \cite{Bellissard1983b, Bellissard1983a,  Sinai1987, Chulaevsky1991, Chulaevsky1993, Eliasson1997, Eliasson2002}. Other diagonalizing flows have been
discussed in a variety of contexts \cite{Deift1983, Brockett1991, Grote2002, Wegner2006},
but perhaps the closest connection to the present work is the similarity renormalization 
group \cite{Glazek1993}.

\section{First Step} \label{2}

\noindent In the first step, we derive an equation for the eigenfunctions of
$H^{(\Lambda)}$.  At this stage, the expansion is just first-order perturbation theory in
$J_0$.  In terms of the $J_0=0$ eigenfunctions $\psi_x^{(0)}(y)=\delta_{xy}$, the state
$\psi_x^{(0)}$ connects to nearest-neighbor states $\psi_y^{(0)}$ with $|x-y|=1$.
Multistep graphs will result when we orthonormalize the new basis vectors.

\subsection{Resonant Blocks} \label{2.1}

\noindent We say that a pair sites $x$ and $y$ are resonant in step 1 if $|v_x -
v_y|<\varepsilon$ and $|x-y|=1$.  Take $\varepsilon=J_0^\delta$ with $\delta=\frac{1}{20}$.    Let

\begin{equation}
S_1=\{x \in  \Lambda : x  \text{ is in a resonant pair}\}. \label{(2.0)}
\end{equation}

\noindent This set can be decomposed into connected components or blocks $B^{(1)}_1,
\ldots, B^{(1)}_N$ based on the graph of resonant links $\langle x,y\rangle$.  Each block is treated as a
model space in quasi-degenerate perturbation theory, so we do not perturb with respect to
couplings within a block.  Our goal for this step is to find a basis in which $H$ is block
diagonal up to terms of order $J_0^2$.    

Let us estimate the probability of $\mathcal{E}_{xy}$, the event that two sites $x,y$ lie
in the same resonant block. If $\mathcal{E}_{xy}$ occurs, then there must be a self-avoiding
walk $\omega$ of resonant links from $x$ to $y$.  We claim that
\begin{equation}
P\left(\mathcal{E}_{xy}\right) \leq \sum\limits_{\omega : x \rightarrow y} \prod\limits_{\langle z,\tilde{z} \rangle\in
\omega} P\big(\langle z,\tilde{z} \rangle \, \text{is resonant}\big) \leq \sum\limits_{\omega : x \rightarrow y}(2 \rho_0
\varepsilon)^{|\omega|} \leq (c_D \rho_0 \varepsilon)^{|x-y|}. \label{(2.1)}
\end{equation}
Because $\omega$ is loop-free, we can change variables replacing $\{v_z\}_{z\in \omega, z
\not= x}$ with $\{v_z - v_{\tilde{z}}\}_{\langle z,\tilde{z} \rangle\in \omega}$ and the Jacobian is $\pm 1$.  Hence the
probability that all the links of $\omega$ are resonant is less than $(2\rho_0 \varepsilon)^{|\omega|}$,
where $|\omega|$ is the number of links in $\omega$.

\subsection {Effective Hamiltonian} \label{2.2}

\noindent Having identified the resonant blocks and having estimated their probabilities,
we proceed to perturb in the nonresonant couplings.  Using the notation
$\langle\psi_x^{(0)}|H|\psi_y^{(0)}\rangle=H_{xy}$, write $H$ as a sum of diagonal and off-diagonal
parts:  $H=H_0 + J$ with
\begin{equation}
H_{0,xy}=H_{xy}\delta_{xy}=v_x \delta_{xy} \equiv E_x \delta_{xy}, \label{(2.7)}
\end{equation}
\begin{equation}
J_{xy}=\begin{cases}-J_0, &|x-y|=1; \\
0, &\text{otherwise}.\end{cases} \label{(2.8)}
\end{equation}
Let us write 
\begin{equation}
J=J^{\per}+J^{\res}, \label{(2.8a)}
\end{equation}
where $J^{\per}$ only contains perturbative
links $\langle x,y \rangle$ with both endpoints outside $S_1$.  Links with at least one of $x,y$ in $S_1$
are in $J^{\res}$ (could be resonant).

First-order Rayleigh-Schr\"odinger perturbation theory would give
\begin{equation}
\psi_{x}^{(1)} = \psi_x^{(0)} + \mathop\sum\limits_{y \not= x}
\frac{\langle\psi_x^{(0)}|J^{\per}|\psi_{y}^{(0)}\rangle\psi_y^{(0)}}{E_{x} - E_{y}} 
=\psi_{x}^{(0)} + \mathop\sum\limits_{y} \frac{J_{xy}^{\per}}{E_{x} - E_{y}}\psi_y^{(0)}.
\label{(2.9)}
\end{equation}
Let us define an antisymmetric operator $A$ with matrix elements
\begin{equation}
A_{xy}=\frac{J^{\per}_{xy}}{E_x - E_y}. \label{(2.10)}
\end{equation}
Then, instead of (\ref{(2.9)}), we use the orthogonal matrix $\Omega=e^{-A}$ for our
basis change:
\begin{equation}
\psi_x^{(1)}=\mathop\sum\limits_{y} \Omega_{xy}^{\tr} \psi_y^{(0)}, \label{(2.11)}
\end{equation}
with $\Omega^{\tr}=e^{A}$ taking the place of $1+A$, which appears in (\ref{(2.9)}).  More
generally, if $J$ were self-adjoint rather than symmetric, then $\Omega$ would be unitary.
A similar construction was used in \cite{Datta1996,Datta1999}.

Let us write $H$ in the new basis:
\begin{equation}
H^{(1)}=\Omega^{\tr}H\Omega=\Omega^{*}H\Omega. \label{(2.12)}
\end{equation}
Then we can define $J_{xy}^{(1)}$ through
\begin{equation}
H_{xy}^{(1)}=E_{x}^{(0)}\delta_{xy} + J_{xy}^{(1)} = H_{0,xy} + J_{xy}^{(1)}. \label{(2.13)}
\end{equation}
The matrix $J^{(1)}$ is no longer strictly off-diagonal.  However, we will see
that $J_{xx}^{(1)}$ is of order $J_0^2 /  \varepsilon$, which is natural since
energies vary only at the second order of perturbation theory when the perturbation is
off-diagonal.  In later stages we will need to adjust $H_0$, but here we may use
$H_0^{(1)}=H_0$.  Observe that $[A, H_0]=-J^{\per}$:
\begin{equation}
[A,H_0]_{xy}=\frac{J^{\per}_{xy}E_y - E_x J^{\per}_{xy}}{E_x - E_y}=-J^{\per}_{xy}. \label{(2.14)}
\end{equation}
Then, using $H=H_0 + J$, we have $[A,H]=-J^{\per}+[A,J]$, and so
\begin{align}
H^{(1)} &= e^{A}He^{-A}=H+[A,H]+\frac{[A, [A,H]]}{2!} + \ldots \notag\\
&= H_0 + J^{\res} + J^{\per} - J^{\per}+[A,J]+\frac{[A, -J^{\per}+[A,J]]}{2!} + \ldots \notag\\
&=H_0+ J^{\res}+\sum\limits_{n=1}^{\infty} \frac{(\ad A)^{n}}{n!}
J-\sum\limits_{n=1}^{\infty} \frac{(\ad A)^{n}}{(n+1)!}J^{\per} \notag\\
&=H_0 + J^{\res} +\sum\limits_{n=1}^{\infty}\frac{n}{(n+1)!} (\ad A)^n J^{\per}
+\sum\limits_{n=1}^{\infty} \frac{(\ad A)^n}{n!} J^{\res} \notag\\
&= H_0+J^{\res}+J^{(1)}.
\label{(2.15)}
\end{align}
Here $\ad A=[A, \cdot]$.

Observe that in the new Hamiltonian, $H_0$ and $J^{\res}$ are still present, but $J^{\per}$ is
gone.  In its place is a series of terms of the form $A^pJ^{\per}A^{q}$ or $A^pJ^{\res}A^{q}$, with $n=p +q\geq 1$.
Since $A_{xy}=J_{xy}^{\per} / (E_x - E_y)$, all such terms are of order $J^{n+1}_{0}
/ \varepsilon^{n}$ with $n \geq 1$.  This means that the new $J^{(1)}_{xy}$ is of
order $J^{2}_{0} / \varepsilon$.  In particular, the matrix elements of $H^{(1)}$ satisfy
\begin{equation}
H_{xy}^{(1)}=E_x \delta_{xy} + J_{xy}^{(1)}=E_x \delta_{xy}+O(J_{0}^{2}/ \varepsilon). \label{(2.16)}
\end{equation}
As in Newton's method, the expansion parameter in each step will be roughly the
square of the previous one.

We would like to interpret the above expressions for $\Omega^{\tr}_{xy}$ and $J_{xy}^{(1)}$
in terms of graphical expansions.  The matrix products $(A^n)_{xy}$ or $(A^{p}JA^{q})_{xy}$
have a natural interpretation in terms of a sum of walks from $x$ to $y$.  At this stage,
$A$ and $J$ allow only nearest neighbor steps.  Thus we may write
\begin{equation}
\Omega^{\tr}_{xy}=\delta_{xy}+\sum\limits_{n=1}^{\infty}
\frac{1}{n!}(A^n)_{xy}=\delta_{xy}+\sum\limits_{G_{1}:x \rightarrow y}
\Omega_{xy}^{\tr}(G_1), \label{(2.17)}
\end{equation}
where $G_1$ is a walk $(x_0 = x, x_1, \ldots, x_n=y)$ with nearest-neighbor steps,
and 
\begin{equation}
\Omega_{xy}^{\tr}(G_1)=\frac{1}{n!} \prod\limits_{p=1}^{n} A_{x_{p-1}x_{p}}. \label{(2.18)}
\end{equation}
In view of the antisymmetry of $A$, the links are oriented, and the walk runs from
$x$ to $y$.  The graphical amplitude obeys a bound
\begin{equation}
|\Omega_{xy}^{\tr}(G_1)|\leq (J_0 / \varepsilon)^{|G_1|}, \label{(2.19)}
\end{equation}
where $|G_1|=n$ denotes the number of steps in $G_1$.

In a similar fashion, we may write
\begin{equation}
J_{xy}^{(1)}=\sum\limits_{g_1:x \rightarrow y} J_{xy}^{(1)}(g_1). \label{(2.20)}
\end{equation}
A graph of $g_1$ consists of $p$ $A$-links, followed by one $J$ link $(J^{\res}$ or
$J^{\per})$, followed by $q$ $A$-links, with $p+q\ge1$.  Thus $g_1$ specifies the following data: a spatial graph,
$g_1^{\sss}$, of unit steps in the lattice, and an assignment of $J^{\res}$ or $J^{\per}$ to one of the steps, with the rest being A-links. If we consider the set of denominators $(E_x - E_y)^{-1}$ associated with the factors $A_{xy}=J_{xy}^{\per} / (E_x - E_y)$, we obtain a denominator graph, $g_1^{\ddd}$.  The amplitude corresponding to $g_1$ is the product of the specified matrix elements and an overall
factor easily derived from (\ref{(2.15)}).  For example, there is a term 
\begin{equation}
(-1)^{n-p} \frac{n}{(n+1)!} \begin{pmatrix}n\\p \end{pmatrix} \prod\limits_{m=1}^{p}
A_{x_{m-1}x_{m}} J^{\per}_{x_{p}x_{p+1}} \prod\limits_{m=p+2}^{n+1} A_{x_{m-1}x_{m}}.
\label{(2.20a)}
\end{equation}
(The binomial coefficient arises from expanding out $(\ad A)^n J^{\per}$ and gathering like terms.) Since the prefactor is bounded by 1, we have an estimate:
\begin{equation}
|J_{xy}^{(1)}(g_1)|\leq J_0(J_0 / \varepsilon)^{|g_1|-1}, \label{(2.20b)}     
\end{equation}
Note that while $J_{xy}^{(1)}(g_1)$ is not symmetric under $x\leftrightarrow y$,
the sum over $g_1$ consistent with a given spatial graph is symmetric (because $A$ is
antisymmetric and $J$ is symmetric).  

\subsection{Small Block Diagonalization}\label{2.3}

We have treated the nonresonant links perturbatively so as to diagonalize the Hamiltonian
up to terms of order $J_0^2$.  In order to finish the first step we get rid of as many of
the remaining $O(J_0)$ terms as possible by diagonalizing within small blocks.  Since there
remains $O(J_0)$ terms connecting the resonant region $S_1$ to its complement, we let
$\overline{S}_1$ be a thickened versions of $S_1$, obtained by adding all first neighbors of $S_1$.  
Then any term in the Hamiltonian with at least one end point in $\overline{S}_1^{\textrm{c}}$ is
necessarily second-order or higher.
Components of ${\overline S}_1$ with volume no greater than $\exp(M2^{2/3})$ will be
considered ``small'' (we take $M=2D$).  For such components, a volume factor $\leq\exp(M2^{2/3})$ will be
harmless in the second step expansion, which has couplings $O(J_0^2)$.  The volume factor
arises from the sum over states in a block.  Small components of ${\overline S}_1$ will be
denoted ${\overline b}_{\alpha}^{(1)}$.

The remaining large components of $\overline{S}_1$ will be denoted
${\overline B}_{\alpha}^{(1')}$, and their union will be denoted ${\overline S}_{1'}$.  If we remove
the one-step collar in each ${\overline B}_{\alpha}^{(1')}$, we obtain $B_{\alpha}^{(1')}=S_1
\cap {\overline B}_{\alpha}^{(1)}$.  Likewise, $S_{1'} = {\overline S}_{1'} \cap S_1$.  In this
way we may keep track of the ``core'' resonant set that produced each large block
${\overline B}_{\alpha}^{(1)}$.   

\begin{figure}[h]
\centering
\includegraphics[width=.75\textwidth]{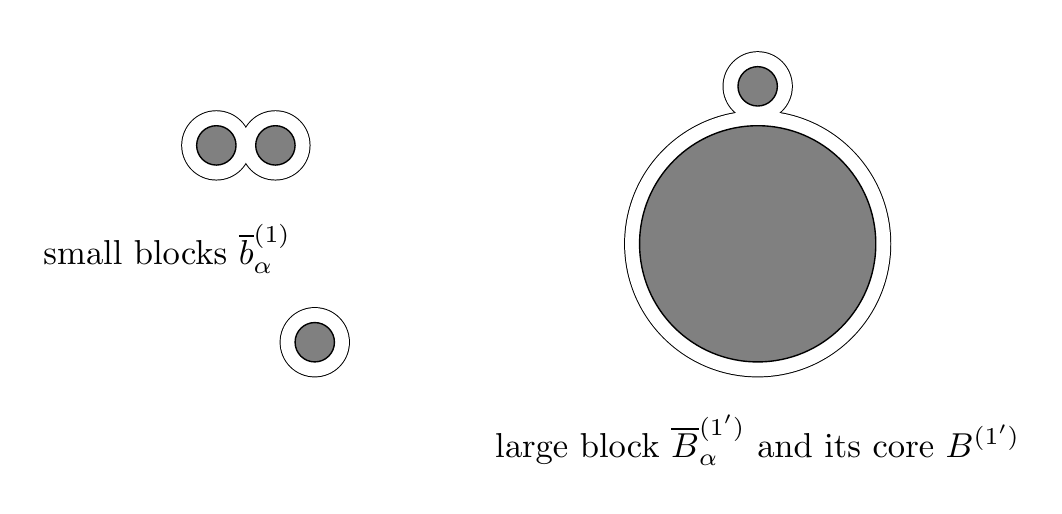}
\caption{The shaded region is $S_1$, the resonant set. $\overline{S}_1$ includes the
collar regions. $S_{1'}$ includes only large blocks $B^{(1')}$, whereas $\overline{S}_{1'}$
is the union of the collared blocks $\overline{B}^{(1')}$. \label{fig1}}
\end{figure}

It is useful to gather terms that are internal to the collared blocks ${\overline b}^{(1)}$
and ${\overline B}^{(1)}$.  
The sum of such terms will be denoted $J_{xy}^{\intint}$; the sum of the remaining terms will be denoted $J_{xy}^{\ext}$. Thus we
write
\begin{equation}\begin{aligned}
J_{xy}^{(1)\intint} &= J_{xy}^{\res} + \sum\limits_{g_1: x \rightarrow y, \, g_1 \cap S_1
\not= \varnothing, \, g_1 \subset {\overline S}_1} J_{xy}^{(1)}(g_1)=J_{xy}^{(1)\sint} +
J_{xy}^{(1)\lint}, \\
J_{xy}^{(1)\ext} &=\sum\limits_{g_1: x\rightarrow y \, \textrm{such that} \, g_1 \cap S_1 = \varnothing \, \textrm{or} \, g_1\not\subset {\overline S}_1}
J_{xy}^{(1)}(g_1), \label{(2.21)}
\end{aligned}
\end{equation}
where $J^{(1)\sint}$ contains terms of $J^{(1)\intint}$ whose graph intersects a small block
${\overline b}^{(1)}$, and $J^{(1)\lint}$ contains terms whose graph intersects a large block
${\overline B}^{(1)}$.  These terms connect a block $b^{(1)}$ or $B^{(1)}$ to its collar.
Then we have
\begin{equation}
H^{(1)}=H_0+J^{(1)\ext}+J^{(1)\sint} + J^{(1)\lint}. \label{(2.21a)}
\end{equation}

We now diagonalize within small blocks ${\overline b}^{(1)}$.  While we can find an
orthogonal matrix $O$ that accomplishes this, we lose control over decay of eigenfunctions
in the block.  Let $O$ diagonalize $H_0 +J^{(1)\sint}$.  
Note that the discriminant of the matrix is analytic in $v$, so it cannot vanish on a set of positive measure without being identically zero. When the $v$'s are separated from each other by $O(J_0)$, it is clear that the eigenvalues are nondegenerate, and so the discriminant is nonzero. Thus the eigenvalues are nondegenerate and the rotation uniquely determined, with probability one.
Note that each block rotation
depends only on $v$ within the block.  Since $H_0+ J^{(1)\sint}$ is block diagonal, $O$ is
also.  Let us define
\begin{align}
H^{(1')} &=O^{\tr} H^{(1)}O=O^{\tr}(H_0 + J^{(1)\ext}+J^{(1)\sint}+J^{(1)\lint})O \notag \\
&=H_0^{(1')}+J^{(1')}+J^{(1)\lint},  \label{(2.22)}
\end{align}
where
\begin{equation}
H_0^{(1')}=O^{\tr}(H_0 +J^{(1)\sint})O \label{(2.22a)}
\end{equation}
is diagonal, and
\begin{equation}
J^{(1')}=O^{\tr}J^{(1)\ext} O=\sum\limits_{g_1} O^{\tr} J^{(1)}(g_1)O. \label{(2.23)}
\end{equation}
Note that the rotation does not affect $J^{(1)\lint}$. Although first-order terms
remain in $J^{(1)\lint}$, large blocks have many resonances and so can be considered
high order after taking the expectation.

Observe from (\ref{(2.20)}) that $J^{(1)}(g_1)$ only has nonzero matrix elements between $x$ and $y$,
where $g_1$ is a walk from $x$ to $y$.  The rotation matrices extend the range of
interaction for $J^{(1)}(g_1)$ to the blocks containing $x$ and $y$.  Let $g_{1'}$ label the
set of terms obtained from the matrix product in (\ref{(2.23)}); it adds at the start and finish of
$g_1$  intra-block jumps associated with matrix elements of $O^{\tr}$ or $O$.  
Thus $g_{1'}$ includes these jumps as additional data; it represents a generalized walk whose first and last steps represent matrix elements of $O$.
Then we may write
\begin{equation}
J_{\alpha \beta}^{(1')}=\mathop\sum_{g_{1'}: \alpha \rightarrow \beta}
J^{(1')}_{\alpha \beta} (g_{1'}) =\sum\limits_{x,y,g_1 : x\rightarrow y}
O^{\tr}_{\alpha x} J_{xy}^{(1)\ext}(g_1)O_{y\beta}. \label{(2.24)}
\end{equation}
Since the matrix elements of $O$ are bounded by 1, (\ref{(2.20b)}) 
leads immediately to the bound
\begin{equation}
|J_{\alpha \beta}^{(1')} (g_{1'})| = |O^{\tr}_{\alpha x} J_{xy}^{(1)\ext}(g_1)O_{y\beta}| \leq J_0(J_0 / \varepsilon)^{|g_{1'}|-1}, \label{(2.25)}
\end{equation}
where $|g_{1'}|=|g_1|$, the length of the walk ignoring intra-block jumps.

Although the eigenfunctions fail to decay in resonant blocks, if we integrate over $v$ we
obtain exponential decay from the probabilities of blocks.  
\begin{proposition}\label{prop:2.1}
Let $\varepsilon = J_0^{1/20}$ be sufficiently small. Then
\begin{equation}
\mathbb{E} \, \sum\limits_{\alpha} |(\Omega O)_{x\alpha}(O^{\tr}\Omega^{\tr})_{\alpha y}|
\leq (c_D^3 \rho_0 \varepsilon)^{|x-y|/3}. \label{(2.26)}
\end{equation}
\end{proposition}
We may think of the rows of $O^{\tr}\Omega^{\tr}$ as the eigenfunctions approximated
to first order, and now including the effects of small blocks.  This is another step towards
proving (\ref{(1.6)}).

\textit{Proof.} Our constructions depend on the collection of resonant blocks, so (\ref{(2.26)}) is best understood
by inserting a partition of unity that specifies the blocks.  Schematically, we may write
\begin{equation}
\mathbb{E} \, \sum\limits_{\alpha} |(\Omega
O)_{x\alpha}(O^{\tr}\Omega^{\tr})_{\alpha y}|=\mathbb{E} \,  \sum\limits_{\mathcal{B}}
\chi_{\mathcal{B}}(v) \sum\limits_{\alpha} |(\Omega O)_{x \alpha}
(O^{\tr}\Omega^{\tr})_{\alpha y}|. \label{(2.27)}
\end{equation}
Here we sum over all possible collections of resonant blocks $\mathcal{B}=\{ \bar{b}^{(1)}_\alpha, \bar{B}^{(1)}_{\alpha '}\}$.  The graphical
expansion (\ref{(2.18)}) for $\Omega$ has to avoid resonant blocks.  We insert it into (\ref{(2.27)}) to
obtain
\begin{equation}
\mathbb{E} \,  \sum\limits_{\mathcal{B}} \chi_{\mathcal{B}}(v) \sum\limits_{G_{1},z, 
\tilde{z}, \tilde{G}_{1}} |\Omega_{xz}(G_1)| \sum\limits_{\alpha}| O_{z\alpha} O^{\tr}_{\alpha
\tilde{z}}| |\Omega_{\tilde{z} y}^{\tr} (\tilde{G}_1)|. \label{(2.28)}
\end{equation}
We bound $\Omega, \Omega^{\tr}$ using (\ref{(2.18)}).  We may also bound
$\Sigma_{\alpha}|O_{z\alpha}O^{\tr}_{\alpha \tilde{z}}|$ by 1 since $O$ is an orthogonal matrix.
Furthermore, if $z \ne \tilde{z}$, and $z, \tilde{z}$ do not belong to the same block, the sum is zero
because the rotations in distinct blocks have non-overlapping supports.  In order for $O_{z\alpha}
O^{\tr}_{\alpha \tilde{z}}$ to be nonzero, $\alpha$ must be both in the block of $z$ and in the block of
$\tilde{z}$.  Thus, in place of $\Sigma_{\alpha} |O_{z\alpha}O_{\alpha \tilde{z}}^{\tr}|$ we may insert an
indicator $\mathbbm{1}_{z\tilde{z}}(v)$ for the event that $z$ and $\tilde{z}$ belong to the same small block
${\overline b}^{(1)}$.  Then (\ref{(2.28)}) becomes 
\begin{equation}
\mathbb{E} \,  \sum\limits_{\mathcal{B}} \chi_{\mathcal{B}}(v)\sum\limits_{G_1, z, \tilde{z},
\tilde{G}_1} (J_0/\varepsilon)^{|G_1|+|\tilde{G}_1|} \mathbbm{1}_{z \tilde{z}}(v)
\leq \sum\limits_{G_1, z, \tilde{z}, \tilde{G}_1} (J_0/\varepsilon)^{|G_1|+|\tilde{G}_1|} \mathbb{E} \,  \mathbbm{1}_{z \tilde{z}}(v), \label{(2.29)}
\end{equation}
where we have interchanged the sum over $\mathcal{B}$ with the sum over $G_1, z, \tilde{z},
\tilde{G}_1$, and used the fact that the sum of $\chi_{\mathcal{B}}(v)$ over $\mathcal{B}$ compatible with
$G_{1},z,\tilde{z}, \tilde{G}_1$ is bounded by 1.

As in (\ref{(2.1)}), the expectation on the right can be bounded by a sum of walks from $z$ to $\tilde{z}$, with resonant
conditions on the links.  We have to allow for the one-step collar in ${\overline b}^{(1)}$, with
nonresonant links.  Still, there must be a (possibly branching) walk from $z$ to $\tilde{z}$ with at
least $\frac{1}{3}$ of the steps resonant.  Thus, we have a probability factor $(\rho_0
\varepsilon)^{1/3}$ for each step of $\omega$ and a factor $J_0 /\varepsilon$ for each step
of $G_1$ and $G_2$.  
The number of branching walks of size $n$ is bounded by $(c_D/4)^n$, where $c_D$ is a constant depending only on the dimension $D$. After summing over $n \ge |x-y|$ and over the choice of factors $(\rho_0\varepsilon)^{1/3}$ or $J_0/\varepsilon$ for each step of the walk, we obtain (\ref{(2.26)}).
%Summing over the combined walk, we obtain (\ref{(2.26)}). 
\qed

\section{The Second Step} \label{3}

\subsection{Resonant Blocks} \label{3.1}

There are some issues that appear for the first time in the second step.  Therefore, it is helpful
to discuss them in the simplest case before proceeding to the general step.

In constructing resonant blocks $B^{(2)}$, we will be allowing links of length 2 or 3 in the
perturbation, which means that it is necessary to check for resonances between states up to 3 steps
apart.  Also, we must consider resonances between states in different blocks $\overline b_{\alpha}^{(1)}$ and
between block states and individual sites.

\textit{Notation and terminology.} Due to the fact that a block state is potentially spread throughout its block, we should consider a
block as a ``supersite'' with multiple states.  The rotation matrix $O$ has one site index and one
state index, see for example (\ref{(2.24)}).  But it would be too cumbersome to maintain a notational
distinction, so we will use $x,y,z$ to denote both sites and states.  If a lattice point $x$ lies
within a block, then the index $x$ may refer to one of the states in ${\overline b}^{(1)}$.  The
labeling of states within a block is arbitrary, so we may choose a one-to-one correspondence between
the sites of ${\overline b}^{(1)}$ and the states of ${\overline b}^{(1)}$, and use that to assign
labels to states.  Block states have many neighbors.  We let $B(x)$ denote the block of $x$. This means that $B(x) = x$ if $x$ is a site; otherwise $B(x)$ is the block that gave rise to the state $x$.

For each $g_{1'}$ corresponding to a term of $J^{(1')}$ with $2 \leq |g_{1'}|\leq 3$, $B(x) \ne B(y)$, $g_{1'} \cap S_2 = \emptyset$, let us define
\begin{equation}
A_{xy}^{(2)\prov} (g_{1'})=\left| \frac{J_{xy}^{(1')}(g_{1'})}{E_{x}^{(1')}-E_{y}^{(1')}}\right|.
\label{(3.1)}
\end{equation}
Here $E_{x}^{(1')}$ denotes a diagonal entry of $H_{0}^{(1')}$.  We call these terms ``provisional''
$A^{(2)}$ terms because not all of them will be small enough to include in $A^{(2)}$.  We only
consider couplings between blocks or sites, never within a block or between a site and itself.
Furthermore, only terms up to third order are considered in this step.  

We say that $g_{1'}$ from $x$ to $y$ is resonant in step 2 if $|g_{1'}|$ is 2 or 3, and if either of
the following conditions hold:
\begin{equation}
\begin{aligned}
&\text{I.}  \quad \big|E_{x}^{(1')} - E_{y}^{(1')}\big| < \varepsilon^{|g_{1'}|}; \\
&\text{II.}  \quad A_{xy}^{(2)\prov}(g_{1'}) > (J_0 / \varepsilon)^{|g_{1'}|} \,\mathrm{ with }\, |x-y|^{(1)} \geq
\tfrac{7}{8} |g_{1'}|. \end{aligned} \label{(3.2)}
\end{equation}
Here $|x-y|^{(1)}$ is the distance from $x$ to $y$ 
in the metric where blocks $\overline{b}^{(1)}$ are contracted to points.
Condition II graphs are nearly self-avoiding, which allows for good Markov
inequality estimates.  Graphs with $|x-y|^{(1)} < \frac{7}{8}|g_{1'}|$ do not reach as far, so
less decay is needed, and we can rely more on inductive estimates.

The graphs $g_{1'}$ that contribute to $A_{xy}^{(2)\prov}$ have the structure 
$AJ$, $JA$, $AAJ$, $AJA$, $JAA$
where $A,J$ are one-step links from the first step.  (Only $J^{\per}$ terms contribute, because
$J^{(1)\sint}$ is gone and $J^{(1)\lint}$ terms connect to some $B^{(1')}$.)  For example, if $ g_{1'}$
specifies  an $AJ^{\per}$ graph between sites $x,y$, then
\begin{equation}
A_{xy}^{(2)\prov}(g_{1'})=\left| \frac{\frac{1}{2} J_{xz}
J_{zy}^{\per}}{\big(E_{x}^{(1')}-E_{y}^{(1')}\big)(E_{z}-E_{y})} \right|=
\frac{\frac{1}{2}J_{0}^{2}}{\big|E_{x}^{(1')}-E_{y}^{(1')}\big||E_z - E_y|}. \label{(3.3)}
\end{equation}
More generally, if either $x$ or $y$ is a block state, then the rotation matrix elements
must be inserted as per (\ref{(2.24)}).  The resonance condition amounts to a condition on products of 2 or
3 energy denominations -- one is $E_{x}^{(1')} - E_{y}^{(1')}$, and the others are specified by
$g_{1'}$.

We have to consider double- and triple-denominator resonant conditions because if, say, one had
$\big|E_{x}^{(1')}-E_{y}^{(1')}\big|\geq \varepsilon^2$ and $|E_x - E_z|\geq \varepsilon$, then the product would
be $\geq \varepsilon^3$.  This is insufficient if one seeks a procedure which can be extended to all
length scales.  A complication with our definition of resonant events is their degree of
correlation.  In the first step, the correlation was mild because probabilities could be estimated
in terms of uncorrelated Lebesgue integrals.  In the present case, we use a Markov inequality
argument to estimate each probability by a product of certain Lebesgue integrals.  Still, if a graph
$g_{1'}$ has returns (loops) or if two graphs overlap at more than one vertex, then the probability
estimate weakens due to correlations.  This makes it harder to control event sums.

As mentioned above, overlapping graphs $g_{1'}$ are problematic for our estimates.  Therefore, we
need to show that there is a sufficient number of non-overlapping graphs to obtain the needed decay
in probabilities.  The following construction generalizes to the $k^\mathrm{th}$ step, so one may imagine the
graphs $g_{1'}$ being arbitrarily long.

Let us define the step 2 resonant blocks.  Consider the collection of all step 2 resonant graphs
$g_{1'}$.  Two resonant graphs are considered to be connected if they have any sites or blocks in
common.  Then the set of sites/blocks that belong to resonant graphs $g_{1'}$ are decomposed into
connected components. 
The result is defined to be the step 2 resonant blocks $B^{(2)}_{1}, \ldots, B_{n}^{(2)}$.  These
blocks do not touch large blocks $B^{(1')}$ because resonant $g_{1'}$ do not.

Note that all sites (states) within a small block ${\overline b}^{(1)}$ are considered to be at a
distance $0$ from each other, hence are automatically connected.  In principle, one could explore
connections between states of a block, but it is impractical because we only know how to vary block
energies as a group; we have no control over intra-block resonances.
Small blocks ${\overline b}^{(1)}$ may be extended or linked together, and there may be entirely new
blocks.  Unlinked small blocks ${\overline b}^{(1)}$ are not held over as scale 2 blocks.

\begin{figure}[h!]
\begin{center}
\includegraphics[width=.7\textwidth]{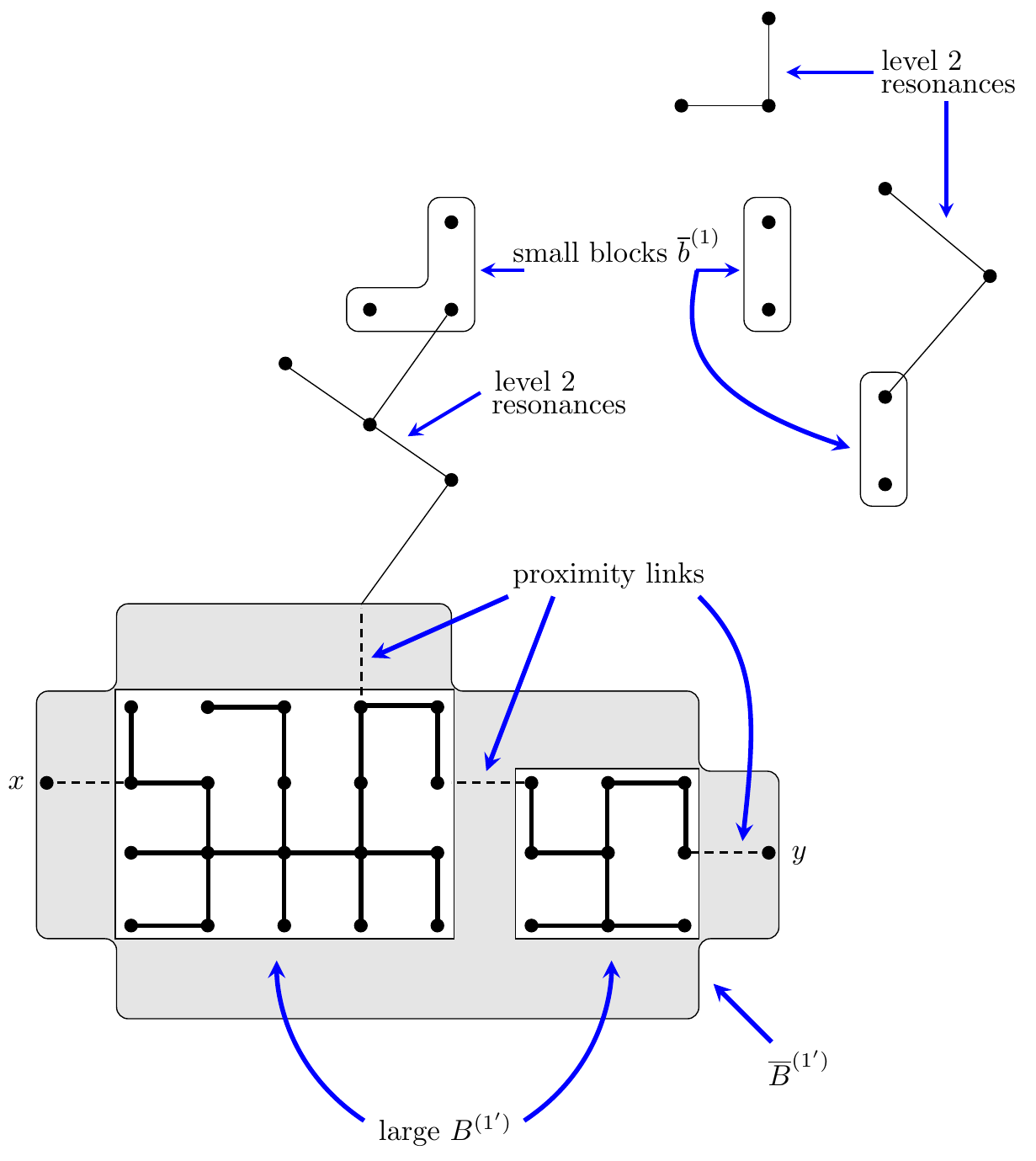}
\caption{Large blocks $B^{(1')}$ and step 2 resonant links  form step 2 blocks $B^{(2)}$. Small blocks $\overline{b}^{(1)}$ without level 2 links are treated perturbatively in this step. Perturbations involving blocks $B^{(2)}$ are deferred to later steps.\label{fig2}}
\end{center}
\end{figure}

Next we add a 3-step collar to all blocks $B^{(2)}$ as well as our leftover large blocks
$B^{(1')}$.  This represents the range of sites reachable by graphs of the order considered in this
step. 
 Since steps may link to small blocks ${\overline b}^{(1)}$, the collar may extend farther than 3
lattice steps, depending on the configuration of small blocks.  We do not expand links involving
blocks $B^{(2)}, B^{(1')}$ at this stage, so the blocks need to expand into the region they could
have linked to.  As in the previous step, we define the resonant region $S_2$ to be the union of the
blocks $B^{(2)}_{\alpha}$ and $B^{(1')}_{\alpha'}$.  Then ${\overline S}_2$ is the collared 
version of $S_2$, and its
components may be divided in to small blocks ${\overline b}^{(2)}_{\alpha}$ (volume $\leq
\exp(M4^{2/3})$) and large blocks ${\overline B}^{(2')}_{\alpha'}$ (volume $> \exp(M4^{2/3})$).  The
union of the ${\overline B}^{(2')}_{\alpha'}$ is denoted ${\overline S}_{2'}$, and then
$B^{(2')}_{\alpha'}\equiv S_2 \cap {\overline B}_{\alpha'}^{(2')}$ and $S_{2'} = {\overline S}_{2'}
\cap S_2$.

We have constructed resonant blocks as connected components of a generalized percolation problem. The following proposition establishes exponential decay of the corresponding connectivity function.
\begin{proposition}\label{prop:3.1}
Let  $\mathcal{E}_{xy}^{(2)}$ denote the probability that $x,y$ lie in the same block ${\overline
b}^{(2)}$ or ${\overline B}^{(2)}$, and let $\varepsilon = J_0^{1/20}$ be sufficiently small.  Then
\begin{equation}
P(\mathcal{E}^{(2)}_{xy}) \leq (c_D^{10} \rho_1 \varepsilon)^{|x-y|/10}. \label{(3.6a)}
\end{equation}
\end{proposition}

%We will bound $\mathcal{E}_{xy}^{(2)}$, the probability that $x,y$ lie in the same block ${\overline{b}^{(2)}$ or ${\overline B}^{(2)}$.  
\textit{Proof.} In the first-step analysis, 
there had to be an unbroken chain
of resonant links from $x$ to $y$.  Here, we need to consider chains formed by ${\overline
B}^{(1')}$ and by resonant graphs $g_{1'}$, each thickened by three steps.  (Let ${\overline
g}_{1'}$ denote the thickened version of $g_{1'}$.)  But when two graphs overlap, we cannot take
the product of their probabilities.  Correlation is manifested by the lack of independent variables
with which to integrate the energy denominators.  To overcome this problem, we find a collection of
non-overlapping ${\overline g}_{1'}, {\overline B}^{(1')}$ which extend at least half the distance
from $x$ to $y$.  In this fashion, we may work with effectively independent events, while giving up
half the decay.

\begin{figure}[h]
\centering
\includegraphics[width=.6\textwidth]{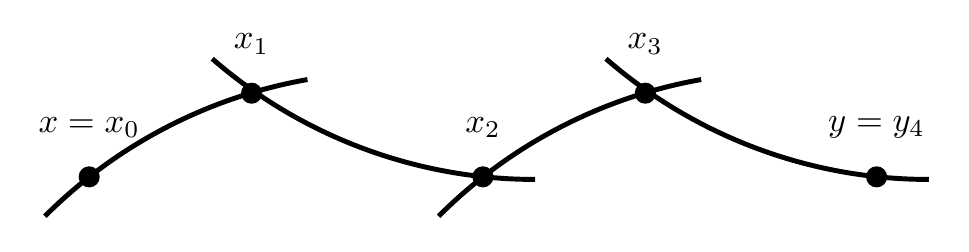}
\caption{Graphs $g_{1',1}$ through $g_{1',4}$ form a bridge from $x$ to $y$. Graph $g_{1',1}$ is disjoint from $g_{1',3}$, and $g_{1',2}$ is disjoint from $g_{1',4}$. \label{fig3}}
\end{figure}

Let $B_{xy}$ be the resonant block containing $x,y$.  Define a metric on $B_{xy}$ by letting $\rho(x_1, x_2)$
be the smallest number of resonant graphs ${\overline g}_{1'}$ or blocks ${\overline B}^{(1')}$
needed to form an unbroken chain from $x_1$ to $x_2$.  If $\rho(x,y)=n$, then there is a sequence of
sites $x=x_0, x_1, \ldots, x_n=y$ such that $\rho(x, x_j)=j$ for $j=0,1, \ldots, n$, with each pair
$\{x_{j-1}, x_j\}$ contained in some resonant graph or block.
Note that the odd-numbered graphs/blocks form a non-overlapping collection of graphs; likewise the
even-numbered ones.  See Figure \ref{fig3}. For if the $j^{\mathrm{th}}$ graph/block overlaps with the $k^{\mathrm{th}}$ one with $k > j+1$, then
one could get from $x$ to $y$ in fewer than $n$ steps.  This construction allows us to bound the
probability of the whole collection of graphs/blocks by the geometric mean of the probabilities of
the even and odd subsequences.  Thus, we may restrict attention to non-overlapping collections of
resonant graphs, losing no more than half the decay distance (from the square root in the geometric
mean).

We have already proven that the probability that $x,y$ belong to the same block ${\overline
B}^{(1')}$ is bounded by $(c_D^3 \rho_0 \varepsilon)^{|x-y|/3}$ -- see the proof of Proposition \ref{prop:2.1}.  Let us focus,
then, on a thickened resonant graph ${\overline g}_{1'}$, and prove an analogous estimate.

Take the simplest case, $|g_{1'}|=2$, condition II of (\ref{(3.2)}), with $A_{xy}^{(2)\prov}$ given by
(\ref{(3.3)}).  Then by the Markov inequality, we have
\begin{align}
P\left(A_{xy}^{(2)\prov} > (J_0 / \varepsilon)^2\right) &\leq \mathbb{E}\left(\big(A_{xy}^{(2)\prov}\big)^s / (J_0 / \varepsilon)^{2s}\right)\notag\\
 &\leq \varepsilon^{2s} \, \mathbb{E} \, \frac{1}{\big|E_{x}^{(1')}-{ E}_{y}^{(1')} \big|^{s} \left|E_z -E_y\right|^s} 
\leq (\rho_1
\varepsilon^s)^2  \label{(3.4)}
\end{align}
for some fixed $s < 1$, say $s=4/5$.  Here $\rho_1$ is a bound for $\sup_{v_0} \int d\lambda(v-v_0) v^{-s}$. This step involves a simple change of variable from
the $v$'s to differences of $v$'s.  By construction, the three sites/blocks $x,y,z$ are distinct, so
the differences are independent.  If $x$ or $y$ is in a block, we make a change of variables to
difference variables on a tree spanning each block ${\overline b}^{(1)}$.  But there is necessarily
one variable left over corresponding to uniform shifts of the potential on that block.  The energies
$E_{x}^{(1')}$ come from diagonalizing $H_0 + J^{\sint}$ on blocks.  They move in sync with the
variables for block shifts of the potential, since each block variable multiplies the identity
operator on its block.  Thus we can use $E_{x}^{(1')}-E_{y}^{(1')}$ and $E_z - E_y$ as integration
variables, and the Jacobian is 1.

If $|g_{1'}|=3$, then an analogous bound
\begin{equation}
P\left(A_{xy}^{(2)\prov} > (J_0/\varepsilon)^3\right) < (\rho_1 \varepsilon^{s})^3 \label{(3.5)}
\end{equation}
holds, provided $g_{1'}$ is ``self-avoiding,'' \textit{i.e.} it has no returns to sites or blocks, which
would lead to fewer than three independent integration variables.  If $g_{1'}$ does have a return,
then we have only condition I to worry about, because $|x-y|^{(1)} < \frac{7}{8} |g_{1'}|$.  The
probability for condition I is easily seen to be $\leq(\rho_1 \varepsilon)^{|g_{1'}|}$, based on the
size of the integration range.  If we consider the bound on $A_{xy}^{(2)}(g_{1'})$ when
$|x-y|^{(1)}=1, |g_{1'}|=3$, it is $J_0^3/\varepsilon^4 = J_0^{3-4\delta} \leq J_0^2$, because one denominator is $\geq
\varepsilon^3$ and the other is $\geq \varepsilon$.  This will be adequate since we only need decay from
$x$ to $y$.

Let us now consider the bound on $P(\mathcal{E}_{xy}^{(2)})$.  In order for $\mathcal{E}_{xy}^{(2)}$ to
occur, there must be a (possibly branching) walk from $x$ to $y$ consisting of 

\begin{enumerate}
\item at most 3 steps at the start and finish of large blocks $B^{(1')}$ and graphs $g_{1'}$.  These
steps have no small factor because of the collars employed in the construction of $B^{(2)}$ and $\overline{g}_{1'}$.
\item Steps in the lattice, not internal to any ${\overline b}^{(1)}$, coming from resonant graphs
$g_{1'}$.  These result in a small probability factor $(\rho_1 \varepsilon^s)^{1/2}$ per
(\ref{(3.4)}), (\ref{(3.5)}), with the square root coming from the geometric mean as discussed above.
\item Steps in lattice which are internal to large blocks ${\overline B}^{(1')}$ or to a ${\overline
b}^{(1)}$ that is part of a resonant graph $g_{1'}$.  These result in a small probability factor $(
\rho_1 \varepsilon)^{1/3}$.  Here we begin with $\rho_0 \varepsilon < \rho_1 \varepsilon  $, which is the probability of
a resonant link at level 1, and add the $\frac{1}{3}$ exponent for the 1-step collar which may be
present about any such link. 
\end{enumerate}

Note that in applying (\ref{(3.4)}), (\ref{(3.5)}), we are constrained to consider only the even- or odd-numbered
graphs $g_{1'}$, because of the potential for shared or looping link variables.  But all of the type
(3) steps can be used because they involve difference variables within each ${\overline b}$.  The
Markov inequality bounds (\ref{(3.4)}), (\ref{(3.5)}) involve differences between block/site variables, so there is
no overlap with the intra-block variables.

\begin{figure}[h]
\centering
\includegraphics[width=.75\textwidth]{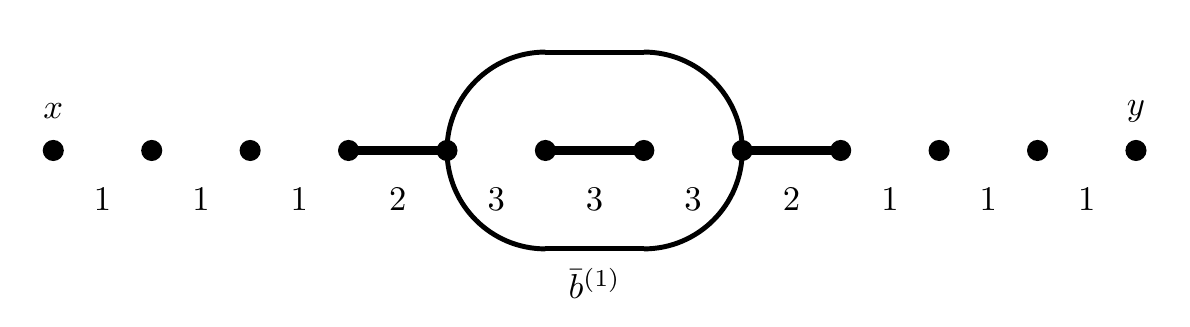}
\caption{The walk from $x$ to $y$ contains six type 1 steps from the collar around $g_{1'}$, two type 2 steps from $g_{1'}$ itself, and three type 3 steps internal to the block $\overline{b}^{(1)}$.  
Type 2 steps are spread out by a factor of 4 to allow for type 1 steps and a factor of 2 to 
account for the geometric mean. Type 3 steps are ``spread out'' by a factor of 3.\label{fig4}}
\end{figure}

For each graph $g_{1'}$, there is a minimum of two type 2 steps and a maximum of 6 type 1 steps, see Figure \ref{fig4}.
Therefore, each small factor from a type 2 step will be spread out over 4 steps by applying an exponent
$\frac{1}{4}$.  Then every step has a factor no worse than $( \rho_1 \varepsilon^s)^{1/8} = (
\rho_1 \varepsilon)^{1/10}$.
Each large block ${\overline B}^{(1')}$ has volume greater than $\exp(M2^{2/3})$, which implies a
diameter greater than $\exp(M2^{2/3}/D)$.  If we take $M=2D$, the diameter is at least 24.  Adding 4
type 1 steps to allow for the collar increase from 1 to 3 on each side, we find that the linear
density of resonant links may decrease from $\frac{1}{3}$ to $\frac{1}{3} \cdot \frac{24}{24+4} =
\frac{2}{7}$.

Combining these facts, we may control the sum over (branching) lattice walks from $x$ to $y$ and over
collections of resonant graphs $g_{1'}$ along the walk as in the proof of Proposition \ref{prop:2.1}. We obtain
\begin{equation}
P(\mathcal{E}^{(2)}_{xy}) \leq (c_D^{10} \rho_1 \varepsilon)^{|x-y|/10}, \label{(3.6)}
\end{equation}
which completes the proof of Proposition \ref{prop:3.1}.

\textit{Remark.}
The sum over $g_{1'}$ containing a particular point is straightforward at this stage,
since it contains no more than 3 steps, and the number of states in a block ${\overline b}^{(1)}$ is
bounded by $\exp(M2^{2/3})$.  We will need to be more careful when we revisit this estimate in the
$k^{\mathrm{th}}$ step. 

\subsection{Perturbation in the Nonresonant Couplings} \label{3.2}

Let us begin by making a split analogous to (\ref{(2.8a)}):
\begin{equation}
J^{(1')}=J^{(1')\per} + J^{(1')\res}. \label{(3.7)}
\end{equation}
Here the perturbation terms are given by
\begin{equation}
J_{xy}^{(1')\per}=\sum\limits_{g_{1'}:x\rightarrow y, \,2 \leq|g_{1'}|\leq 3, \, g_{1'} \cap
S_{2}=\varnothing,\, B(x)\ne B(y)} J_{xy}^{(1')}(g_{1'}), \label{(3.8)}
\end{equation}
and the ``resonant'' part $J_{xy}^{(1')\res}$ consists of all terms $J_{xy}^{(1')}$ with
$|g_{1'}| \geq 4$, terms intersecting $S_2$, and diagonal/intrablock terms for which $B(x)$ (the block of $x$) is the same as $B(y)$.  Let us put 
\begin{equation}
A_{xy}^{(2)}  =
\sum\limits_{g_{1'}} A_{xy}^{(2)}(g_{1'}) =\frac{J_{xy}^{(1')\per}(g_{1'})}{E_{x}^{(1')}-E_{y}^{(1')}}. \label{(3.9)}
\end{equation}
We would like to ``resum'' all terms from long graphs $g_{1'}$
from $x$ to $y$, \textit{i.e.} those with $|g_{1'}| > \frac{8}{7}|x-y|^{(1)}$.  These are small enough,
uniformly in $v$, so there is no need to keep track of individual graphs and their $v$-dependence.
Let $g_{1''}$ denote either a short graph from $x$ to $y$ or a special jump step from $x$ to $y$
whose length is defined to be $2$.  The jump step represents the collection of all long graphs from
$x$ to $y$.  We call $g_{1''}$ short or long accordingly.  Then put
\begin{equation}
A_{xy}^{(2)}(g_{1''})=\begin{cases}A_{xy}^{(2)}(g_{1''}), \ \text{if} \ g_{1''}=g_{1'}, 
 \text{a short graph};\\ \sum\limits_{\mathrm{long} \, g_{1'}: x \rightarrow y} 
A_{xy}^{(2)}(g_{1'}), \ \text{if} \ 
g_{1''} \,\text{is long}. \end{cases} \label{(3.10)}
\end{equation}

Now define the basis-change operator 
\begin{equation}
\Omega^{(2)}=e^{-A^{(2)}}, \label{(3.11)}
\end{equation}
and the new Hamiltonian 
\begin{equation}
H^{(2)}=\Omega^{(2)\tr} H^{(1')}\Omega^{(2)}. \label{(3.12)}
\end{equation}
Recall that $H^{(1')}= H_0^{(1')}+J^{(1')} + J^{(1)\lint}$ with $H_{0,xy}^{(1')} =
E_{x}^{(1')}\delta_{xy}$.  As in (\ref{(2.15)}), we have
\begin{align}
H^{(2)}&=H_{0}^{(1')}+J^{(1')\res} + J^{(1)\lint} + \sum\limits_{n=1}^{\infty}
\frac{n}{(n+1)!} \frac{(\ad A^{(2)})^{n}}{n!} J^{(1')\per} +\sum\limits_{n=1}^{\infty}
\frac{(\ad A^{(2)})^n}{n!} J^{(1')\res} \notag\\
&= H_{0}^{(1')} + J^{(1')\res} + J^{(1)\lint} + J^{(2)}. \label{(3.13)}
\end{align} 
Note that since $J^{(1')}$ is second order in $J_0$, all commutator terms are fourth-order
or higher.  

Let us describe the graphical expansions for $\Omega^{(2)}$ and $J^{(2)}$.  As in (\ref{(2.17)}), we may
write
\begin{equation}
\Omega_{xy}^{(2)\tr} = \delta_{xy} + \sum\limits_{n=1}^{\infty} \frac{1}{n!}
(A^{(2)n})_{xy}=\delta_{xy} + \sum\limits_{G_{2}: x \rightarrow y} \Omega_{xy}^{(2)\tr}(G_2),
\label{(3.14)}
\end{equation}
where $G_2$ is a walk $(x_0=x, x_1, \ldots, x_n=y)$ consisting of a sequence of $n$
subgraphs $g_{1'',p}:x_{p-1} \rightarrow x_{p}$.  Then
\begin{equation}
\Omega_{xy}^{(2)\tr} (G_2)=\frac{1}{n!} \prod\limits_{p=1}^{n}
A_{x_{p-1}x_{p}}(g_{1'',p}). \label{(3.15)}
\end{equation}
Note that if we put together all the subgraphs of $G_2$, we get a walk $G_2^{\sss}$ from $x$ to $y$ consisting of unit steps of short $g_{1'}$ graphs and jump steps from long $g_{1''}$, in both cases between sites or blocks ${\overline b}^{(1)}$.  In addition, there are jump steps within blocks ${\overline b}^{(1)}$ from rotation matrix elements.  
The length $|G_2|$ is the sum of the
constituent lengths $|g_{1'', p}|$ -- and as explained after (\ref{(2.25)}), these do not include the
intra-block jumps since there is no decay within blocks.  Define the length $|G_2^{\sss}|$ of the spatial graph as the sum of the constituent lengths ignoring long graphs $g_{1''}$.  The graph $G_2$ also determines a graph $G_2^{\ddd}$ of energy denominators.  Each short $g_{1'', p}$ has one or two $A^{(1)}$ factors (each with an energy denominator with 
an $\varepsilon$ cutoff) and an overall energy
denominator with cutoff $\varepsilon^{|g_{1''}|}$.  The number of denominators always equals the
number of steps because each $J^{(1')}_{xy}(g_{1'})$ is always short one denominator.  Hence
$|G_2^{\sss}|=|G_2^{\ddd}|$.  (Long graphs $g_{1''}$ are ignored on both sides of this equality.)

Nonresonant conditions (\ref{(3.2)}) apply for $J_{xy}^{(1')\per}$ links, so each $A^{(2)}_{xy}(g_{1''})$
in (\ref{(3.15)}) is bounded by $(J_0/\varepsilon)^{|g_{1''}|}$.  As explained after (\ref{(3.5)}), the bound for a long
$g_{1'}$ is $J_0^3/\varepsilon^4$.  Allowing a constant for the number of long graphs $g_{1''}$ from
$x$ to $y$, we find that $|A_{xy}^{(2)}(g_{1''})|\leq J_0^2 \leq (J_0 / \varepsilon)^{|g_{1''}|}$ for
long $g_{1''}$.  Hence
\begin{equation}
|\Omega_{xy}^{(2)\tr}(G_2)|\leq \frac{1}{n!} (J_0 / \varepsilon)^{|G_2|}. \label{(3.16)}
\end{equation}

We continue with a graphical representation for $J^{(2)}$, derived from (\ref{(3.13)}):
\begin{equation}
J_{xy}^{(2)}=\sum\limits_{g_{2}:x \rightarrow y}J_{xy}^{(2)}(g_2). \label{(3.17)}
\end{equation}
Here $g_2$ is a generalized walk from $x$ to $y$ with a structure similar to that of
$G_2$.  It consists of $p \ A^{(2)}$-links, followed by one $J^{(1')}$-link $(J^{(1')\per}$ or
$J^{(1')\res}$), followed by $q \ A^{(2)}$-links, with $p + q \geq 1$.  
As above, $A^{(2)}$-links are
short or long (resummed) and indexed by $g_{1''}$; $J^{(1')}$-links are unchanged, indexed by $g_{1'}$.
Note that in the case of $J^{(1')\res}$, the subgraph can have length $|g_{1''}|\geq 4$.  Thus
$J_{xy}^{(2)}(g_2)$ has an expression of the form (in the case of $J^{(1')\per}$)
\begin{equation}
(-1)^{n-p} \frac{n}{(n+1)!} \begin{pmatrix}n\\ p \end{pmatrix} \prod\limits_{m =1}^{p}
A^{(2)}_{x_{m-1}x_{m}} (g_{1'', m}) J^{(1')\per}_{x_{p} x_{p+1}} (g_{1', p+1})
\prod\limits_{m=p+2}^{n+1} A^{(2)}_{x_{m-1}x_{m}} (g_{1'', m}). \label{(3.18)}
\end{equation}
As in the case of $G_2^{\sss}$, the spatial graph $g_2^{\sss}$ formed by uniting all the subgraphs
forms a walk of unit steps and jump steps between blocks/sites, and jump steps within blocks.  The
denominator graph $g_2^{\ddd}$ is short one link, compared with the non-jump steps of $g_2^{\sss}$.  Note
that since $A^{(2)}$ and $J^{(1')}$ are both second-order or higher, all terms $J_{xy}^{(2)}(g_2)$
are of degree at least 4.  If we apply nonresonant conditions (\ref{(3.2)}) to $A^{(2)}$ and previous
bounds (\ref{(2.25)}) on $J^{(1')}$, we see that 
\begin{equation}
|J_{xy}^{(2)} (g_2) | \leq J_0(J_0 / \varepsilon)^{|g_{2}|-1}.  \label{(3.19)}
\end{equation}

\subsection{Small Block Diagonalization} \label{3.3}

In the last section we divided the current singular region ${\overline S}_2$ into large blocks
${\overline B}^{(2)}$ and small blocks ${\overline b}^{(2)}$ with volume bounded by
$\exp(M4^{2/3})$.  By construction, any term of the Hamiltonian whose graph does not intersect $S_2$
is fourth-order or higher.  Put
\begin{align}
J^{(2)}+J^{(1')\res} + J^{(1)\lint} &= J^{(2)\ext} + J^{(2)\intint} \notag\\
&= J^{(2)\ext} + J^{(2)\sint} + J^{(2)\lint}.  \label{(3.20)}
\end{align}
Here $J^{(2)\intint}$ contains terms whose graph intersects $S_2$ and is contained in
${\overline S}_2$.  Let $J^{(2)\lint}$ include terms of $J^{(2)\intint}$ that are contained in large
blocks ${\overline B}^{(2)}$.  Let $J^{(2)\sint}$ include terms of $J^{(2)\intint}$ that are contained
in small blocks ${\overline b}^{(2)}$, as well as second- or third-order diagonal/intrablock terms for sites/blocks in $S_2^{\textrm{c}}$.
All remaining terms of $J^{(2)}$ and $J^{(1')\res}$ are included in $J^{(2)\ext}$.  (There are some
terms fourth-order or higher in $J^{(1')\res}$ that are now in $J^{(2)\ext}$ -- these were not
expanded in (\ref{(3.7)})-(\ref{(3.13)}) since they were already of sufficiently high order, and had too great a range.)

Let $O^{(2)}$ be the matrix that diagonalizes $H^{(1')}_0 + J^{(2)\sint}$.  It acts nontrivially
only within small blocks.  
This includes  blocks $b_\alpha^{(1)}$, which need to be ``rediagonalized'' due to the presence of intrablock interactions of second and third order.
Then put
\begin{align}
H^{(2')} &=O^{(2)\tr}H^{(2)}O^{(2)}\notag\\
&= O^{(2)\tr}\big(H^{(1'')}_0 + J^{(2)\ext} +J^{(2)\sint}+J^{(2)\lint}\big)O^{(2)} \notag\\
&= H_0^{(2')} + J^{(2')} + J^{(2)\lint} , \label{(3.21)}
\end{align}
where
\begin{equation}
H_0^{(2')}=O^{(2)\tr}\big(H_0^{(1')}+ J^{(2)\sint}\big)O^{(2)} \label{(3.22)}
\end{equation}
is diagonal, and
\begin{equation}
J^{(2')}=O^{(2)\tr} J^{(2)\ext} O^{(2)}. \label{(3.23)}
\end{equation}
Note that $J^{(2)\lint}$ is not affected by the rotation.

Recall the graphical expansions (\ref{(3.17)}), (\ref{(2.24)}), which define the terms from $J^{(2)}, J^{(1')\res}$
that contribute to $J^{(2)\ext}$.  These are combined and rotated to produce an analogous graphical
expansion for $J^{(2')}$:
\begin{equation}
J^{(2')}_{\alpha{\beta}} = \sum\limits_{g_{2'}:\alpha \rightarrow \beta}
J^{(2')}_{\alpha{\beta}} (g_{2'}), \label{(3.24)}
\end{equation}
where $g_{2'}$ specifies $x,y$, a rotation matrix element $O^{(2)\tr}_{\alpha x}$,
followed by a graph $g_2$ or $g_{1'}$, and then another rotation matrix element $O^{(2)}_{y\beta}$. 

We complete the analysis of the second step by proving an exponential localization estimate for the
rotations so far.  Let us introduce cumulative rotation matrices:
\begin{equation}
\begin{aligned}
R^{(1)} &= \Omega^{(1)},\\
R^{(1')} &= \Omega^{(1)} O^{(1)},\\
R^{(2)} &= R^{(1')} \Omega^{(2)},\\
R^{(2')} &= R^{(2)} O^{(2)}. \end{aligned} \label{(3.25)}
\end{equation}
\begin{proposition}\label{prop:3.2}
Let $\varepsilon = J_0^{1/20}$ be sufficiently small. Then
\begin{equation}
\mathbb{E} \, \sum\limits_{\alpha} \big|R^{(2')}_{x \alpha} R^{(2')\tr}_{\alpha y}\big|\leq
(c_{D}^{10} \rho_1 \varepsilon)^{|x-y|/10}. \label{(3.27)}
\end{equation}
\end{proposition}

\textit{Proof.} Let us write
\begin{equation}
\mathbb{E} \, \sum\limits_{\alpha} \big|R_{x\alpha}^{(2')} R_{\alpha y}^{(2')\tr} \big| 
=\mathbb{E} \, \sum\limits_{\alpha} \big|(\Omega^{(1)} O^{(1)}\Omega^{(2)}O^{(2)})_{x \alpha}
(O^{(2)\tr} \Omega^{(2)\tr} O^{(1)\tr} \Omega^{(1)\tr})_{\alpha y}\big|, \label{(3.26)}
\end{equation}
introducing as before a partition of unity for collections of blocks, and graphical
expansions for each $\Omega$ or $\Omega^{\tr}$ matrix.  The graphs combine to form a walk from $x$
to $y$, with possible intra-block jumps. 

We need to review how the rotations $O = O^{(1)}$ and $O^{(2)}$ fit together. 
%due to $(O^{(1)}O^{(2)}O^{(2)\tr} O^{(1)\tr})_{z\tilde{z}}$ and to $(O^{(1)}O^{(1)\tr})_{z\tilde{z}}$.
By construction, $\Omega^{(2)}$ is the identity plus a sum of terms involving products of $O^{(1)\tr}_{\alpha x} J_{xy}^{(1)\ext}(g_1)O^{(1)}_{y\beta}$, along with energy denominators -- \textit{c.f.} (\ref{(2.25)}), (\ref{(3.9)}), (\ref{(3.15)}). Thus every $O^{(1)}$ matrix element is followed by a $O^{(1)\tr}$ matrix element -- except for the last one, which is followed by an $O^{(2)}$ matrix element. This structure follows naturally from the process of sequential basis changes. We need to distinguish between the identity matrix term in $\Omega^{(2)}$ and the nontrivial terms. If we have the identity matrix, then we may form the matrix product $\tilde{O} = O^{(1)}O^{(2)}$ before taking absolute values; then we may bound $\sum_\alpha |\tilde{O}_{z\alpha}\tilde{O}^{\tr}_{\alpha \tilde{z}}|$ by 1 as before. For the nontrivial terms, 
the sum over states in the $b^{(1)}$ blocks as well as the sum over their sites are
%the presence of energy denominators forces us to take absolute values before taking the matrix product, so we have to consider each term in the matrix product separately, and this leads to an extra sum over states in the block. 
compensated by the smallness of $A^{(2)}_{xy}(g_{1'})$, which must be present in this case. 
(We could also have taken advantage of the inequality $\sum_\alpha |O^{(1)}_{z\alpha}O^{(1)\tr}_{\alpha \tilde{z}}| \le 1$ at the intermediate blocks $b^{(1)}$; either way those sums are under control.)
In step 2, the size of blocks is limited, so the state sums are of little consequence. But in the $k^{\text{th}}$ step we need to ensure that all state/site sums at blocks match up with appropriately small factors.

In all cases, intrablock jumps
between two sites $z$, $\tilde{z}$ are controlled by the probability that $z$, $\tilde{z}$ belong to the same block
${\overline b}^{(2)}$ or ${\overline b}^{(1)}$.  Each of these probabilities has been estimated
already in terms of a sum over walks from $z$ to $\tilde{z}$ with their associated small probability
factors, \textit{c.f.} the proofs of (\ref{(2.1)}), (\ref{(3.6)}).  Therefore, we can bound (\ref{(3.26)}) in terms of a sum over walks
from $x$ to $y$ with at least a factor of $(\rho_1 \varepsilon)^{1/10} $ per
step.  The bound (\ref{(3.27)}) then follows. \qed

\textit{Remark.} The decay in the nonresonant region is faster, with a rate constant $J_0 /
\varepsilon$, but in this integrated bound, we have to use the larger rate constant associated with
resonant regions.    

\section{The General Step} \label{4}

Let us consider the $k^{\mathrm{th}}$ step of the procedure, focusing on uniformity in $k$.
We work on a length scale $L_k \equiv (\frac{15}{8})^k$, which is slightly smaller than
the na\"{i}ve $2^k$ scaling of Newton's method.  When long graphs are resummed, they are ``contracted'' to $\frac{7}{8}$
of their original length, and this reduces the length scale that can be treated in the following
steps.

\subsection{Starting Point} \label{4.1}

\begin{figure}[h]
\centering
\includegraphics[width=.7\textwidth]{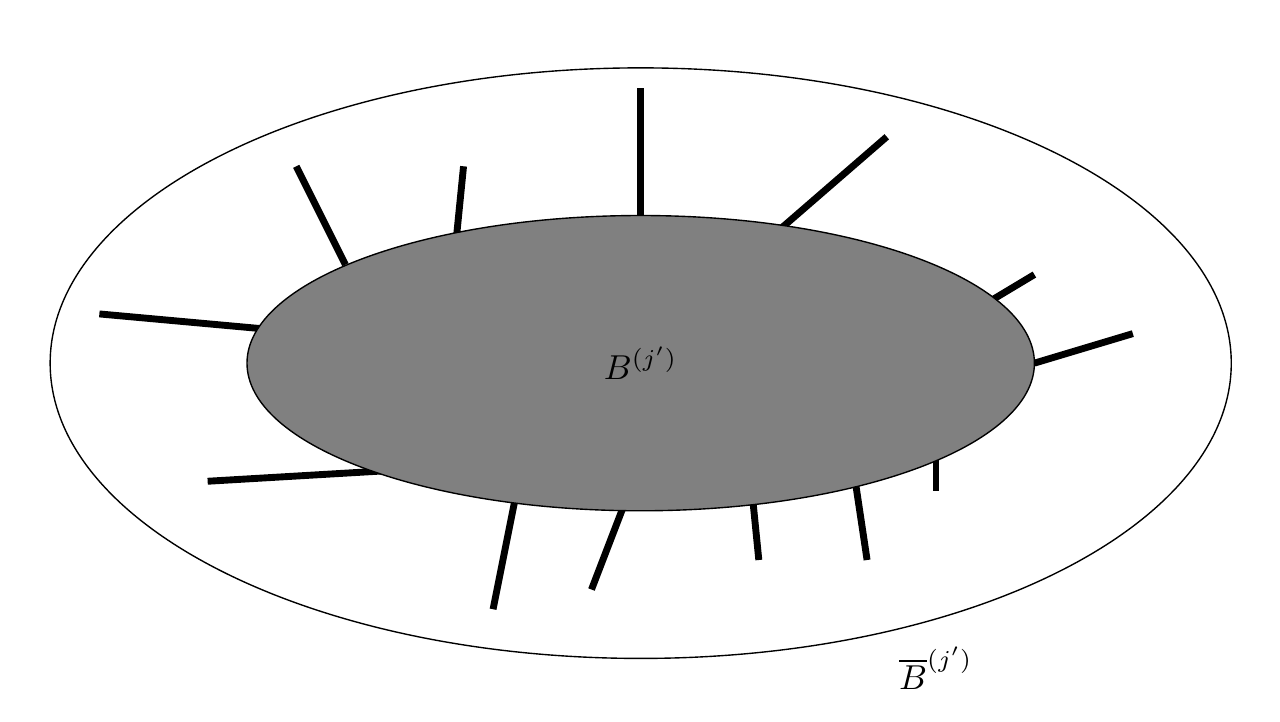}
\caption{Leftover interaction terms couple the core $B^{(j')}$ to its collar.\label{fig5}}
\end{figure}

After $j$ steps, the structure of the resonant regions extends the picture in Figure \ref{fig1}.  Small
blocks ${\overline b}^{(j)}$ have size up to $\exp(ML_j^{2/3})$, which is manageable because
couplings in $J^{(j')}$ are at least $O(J_0^{L_{j}})$ at the conclusion of the $j^{\mathrm{th}}$ step.
Rotations have been performed in small blocks ${\overline b}^{(j)}$, diagonalizing the Hamiltonian
there up to terms of order $L_j$.  Rotations are deferred for large blocks ${\overline B}^{(j')}$
until $j$ is large enough for the volume conditions to be satisfied, at which point they may become
small blocks.  Collar neighborhoods of width $L_j -1$ are added to large blocks $B^{(j)}$ to form blocks ${\overline B}^{(j')}$.  This is so that no couplings to $B^{(j)}$ are
involved in the $j^{\mathrm{th}}$ step expansion (the distance to ${\overline B}^{(j')c}$ is greater than $L_j
- 1$, the maximum order for the $j^{\mathrm{th}}$ step).  In effect, the large blocks $B^{(j')}$ grow ``hair'' -- unresolved interaction terms of length up to $L_j -1$ which could not be expanded because they
are not small enough to beat the volume of the block.  See Figure \ref{fig5}.
We maintain a uniform bound as in (\ref{(3.6)}) on $P(\calE_{xy}^{(j)})$, the probability that
$x,y$ belong to the same large block ${\overline B}^{(j')}$ or small block ${\overline b}^{(j)}$.

Resonances treated in step $j$ involve graphs with $L_{j-1} \leq |g| < L_j$, so the associated
probabilities are of that order in $\varepsilon^s$.  Nonresonant couplings in step $j$ are of the
same order in $J/\varepsilon$. Couplings and resonance probabilities decrease in tandem as $j$
increases, a key feature of our procedure.  

Let us recapitulate the transformations from the $j^{\mathrm{th}}$ step.  After defining resonant blocks
$B^{(j)}$, the Hamiltonian was rewritten as follows:
\begin{align}
H^{(j-1)'} &= H_0^{(j-1)'} + J^{(j-1)'} + J^{(j-1)\intint}\notag\\
&= H_0^{(j-1)'} +J^{(j-1)'\per} + J^{(j-1)'\res}+J^{(j-1)\lint} . \label{(4.1)}
\end{align}
The terms in $J^{(j-1)'\per}$ were ``rotated away'' by conjugating $H^{(j-1)'}$ with
$\Omega^{(j)}$.  This led to a new Hamiltonian with smaller interactions and after regrouping
terms, it became
\begin{align}
H^{(j)} &= H_0^{(j-1)'} + J^{(j)} + J^{(j-1)'\res} +J^{(j-1)\lint}\notag\\
&= H_0^{(j-1)'} +J^{(j)\ext} +J^{(j)\sint} +J^{(j)\lint}. \label{(4.2)}
\end{align}
Rotations $O^{(j)}$ were performed in small blocks, and low-order diagonal terms were
absorbed into $H_0^{(j)}$, leading to a form like the one we started with:
\begin{equation}
H^{(j')}=H_0^{(j')} + J^{(j')} + J^{(j)\lint}. \label{(4.3)}
\end{equation}

The rotation $\Omega^{(j)}$ has a graphical expansion
\begin{equation}
\Omega_{xy}^{(j)\tr} = \delta_{xy} + \sum\limits_{G_{j}: x \rightarrow y}
\Omega_{xy}^{(j)\tr} (G_j) \label{(4.4)}
\end{equation}
as in (\ref{(3.14)}), and as in (\ref{(3.16)}) we have a uniform bound
\begin{equation}
|\Omega_{xy}^{(j)\tr} (G_j)| \leq \frac{1}{n!} (J_0/\varepsilon)^{|G_{j}|}. \label{(4.5)}
\end{equation}
This arises from the more basic estimate:
\begin{equation}
\big|A_{xy}^{(j)} (g_{(j-1)''})\big| \leq \begin{cases}(J_0/\varepsilon)^{|g_{(j-1)''}|},\, \text{in general};\\
J_0^{|g_{(j-1)''}|}, \, \text{if} \ g_{(j-1)''} \ \text{is a jump step}.\ \end{cases}. \label{(4.5a)}
\end{equation}
Here we define for any $i$ and any jump step $g_{i''}$ from $x$ to $y$:
\begin{equation}
|g_{i''}|=|x-y|^{(i)} \lor \tfrac{7}{8} L_{i+1}. \label{(4.5b)}
\end{equation}
Here $ |x-y|^{(i)}$ is the distance from $x$ to $y$ in the metric where blocks $\overline{b}^{(\tilde{i})}$ on scales $\tilde{i}\le i$ are contracted to points.
Likewise, the interaction terms
$J^{(j')}$ and $J^{(j)\lint}$ have graphical expansions generalizing (\ref{(3.24)}).  Thus
\begin{equation}
J^{(j')}_{\alpha \beta} = \sum\limits_{g_{j'}:\alpha \rightarrow \beta} J^{(j')}_{\alpha \beta}
(g_{j'}), \label{(4.6)}
\end{equation}
with bounds as in (\ref{(3.19)}):
\begin{equation}
\big|J^{(j')}_{\alpha \beta} (g_{j'})\big| \leq J_0(J_0 / \varepsilon)^{|g_{j'}|-1}. \label{(4.7)}
\end{equation}

The graphs $G_j$ and $g_{j'}$ are actually ``walks of walks,'' with each step representing a walk
from the previous scale.  When unwrapped to the first scale, we obtain spatial graphs $G_j^{\sss}$ and
$g_{j'}^{\sss}$, as well as denominator graphs $G_j^{\ddd}$ and $g_{j'}^{\ddd}$.  Resummed sections
appear as jump steps with no denominators.  Likewise, rotation matrix elements appear as jump steps
within blocks.

The goal of each step is to prove a bound analogous to (\ref{(3.27)}):
\begin{equation}
\mathbb{E} \, \sum\limits_{\alpha} \big|R^{(j')}_{x \alpha} R^{(j')\tr}_{\alpha y}\big| \leq
(c_D^{50} \rho_1 \varepsilon)^{|x-y|/50}, \label{(4.8)}
\end{equation}
where
\begin{equation}
R^{(j')} = R^{(j)}O^{(j)} = R^{(j-1)'} \Omega^{(j)} O^{(j)} \label{(4.9)}
\end{equation}
is the cumulative rotation matrix, whose columns represent the eigenfunctions
approximated up to scale $L_j$. 

\subsection{Resonant Blocks} \label{4.2}

Here we make straightforward generalizations of definitions from step 2.  Let $g_{j'}$ be a graph
that does not intersect any $B^{(j')}$, with $L_k \leq |g_{j'}| < L_{k+1}$ (recall that $k=j+1$).
Define
\begin{equation}
A_{xy}^{(k)\prov} (g_{j'})=\left| \frac{\tilde{J}_{xy}^{(j')}(g_{j'})}{E_{x}^{(j')} - E_{y}^{(j')}} \right|,
\label{(4.10)}
\end{equation}
where $\tilde{J}^{(j')}_{xy} (g_{j'})$ is the same as $J^{(j')}_{xy} (g_{j'})$, except jump steps $g_{i''}$ that are subgraphs of $g_{j'}$ are replaced with their upper bound $J_0^{|g_{i''}|}$ from (\ref{(4.5a)}). 
We say that $g_{j'}$ from $x$ to $y$ is resonant in step $k$ if either of the following conditions
hold:
\begin{equation}
\begin{aligned}
\mathrm{I.} \  &\big|E_{x}^{(j')} - E_{y}^{(j')}\big| < \varepsilon^{|g_{j'}|};\\
\mathrm{II.}\   &A_{xy}^{(k)\prov} (g_{j'}) > (J_0 / \varepsilon)^{|g_{j'}|}\ \text{with} \ |x-y|^{(j)} \ge
\tfrac{7}{8} |g_{j'}|. \end{aligned} \label{(4.11)}
\end{equation}

It may be helpful to explain the key ideas behind maintaining uniform exponential decay in our constructions. A resonant graph can be thought of as an event with a small probability. In order for a collection of graphs to be rare, we need to be able to sum the probabilities. In the ideal situation, where there are no repeated sites/blocks in the graph, the probability is exponentially small, so it can easily be summed. However, when graphs return to previously visited sites, dependence between denominators develops, and then the Markov inequality that is used to estimate probabilities begins to break down. Subgraphs in a neighborhood of sites with multiple visits need to be ``erased," meaning that inductive bounds are used, and they do not participate in the Markov inequality. (By this we mean that the bound $P(AC > B\overline{C}) \leq \mathbb{E}(A\overline{C})/ (B \overline{C}) =\mathbb{E}(A)/B$ is used when $\overline{C}$ is bound for $C$ -- so the variation of $C$ is not helping the bound.) When there are a lot of return visits, a graph's length $|g_{j'}|$ is shortened by at least a factor $\frac{7}{8}$, and it goes into a jump step, where again we use inductive bounds. In this case, we have more factors of $J_0$, and hence a more rapid decay, and this provides the needed boost to preserve the uniformity of decay in the induction. (Fractional moments of denominators are finite, no matter the scale, which provides uniformity for ``straight" graphs with few returns.) The net result is uniform probability decay, provided we do not sum over unnecessary structure, \textit{i.e.} the substructure of jump steps. Note that jump steps represent sums of long graphs, so when taking absolute values it is best to do it term by term. This is why we replaced jump steps with their upper bound in (\ref{(4.10)}). (The jump step bound (\ref{(4.5a)}) is also a bound on the sum of the absolute values of the contributing graphs.)

Now consider the collection of all step $k$ resonant graphs $g_{j'}$.  Any two such graphs are
considered to be connected if they have any sites or blocks in common. Decompose the set
of sites/blocks that belong to resonant graphs into
connected components.  The result is defined to be the set of step $k$ resonant blocks
$B_{\alpha}^{(k)}$.  By construction, these blocks do not touch large blocks $B^{(j')}$.  Small
blocks ${\overline b}^{(1)}, \ldots, {\overline b}^{(j)}$ can become absorbed into blocks
$B^{(k)}$, but only if they are part of a resonant graph $g_{j'}$.  

A collar of width $L_{k+1}$ must be added to all blocks $B^{(j')}$ and $B^{(k)}$ because we will
not be expanding graphs of that length that touch any of those blocks.  Let $S_k$ be the union of
the blocks $B^{(k)}$ and $B^{(j')}$.  Then ${\overline S}_k$ is the collared version of $S_k$, and
its components are divided into small blocks ${\overline b}_{\alpha}^{(k)}$ (volume $\leq
\exp(ML_{k}^{2/3}))$ and large blocks ${\overline B}_{\alpha}^{(k)}$ (volume $>
\exp(ML_{k}^{2/3}))$.  The union of the ${\overline B}_{\alpha}^{(k)}$ is denoted ${\overline S}_{k'}$,
and then $B_{\alpha}^{(k')}\equiv S_k \cap {\overline B}_{\alpha}^{(k')}$ and $S_{k'} = \overline S_{k'} \cap 
S_k$.

As discussed in Section \ref{3.1}, if $x,y$ belong to the same resonant block $B^{(k)}$, then there must
be a sequence of resonant graphs connecting $x$ to $y$ with the property that the even and odd
subsequences consist of non-overlapping graphs.  This allows us to focus on estimating probabilities
associated with individual resonant graphs.

The next three subsections establish key results that will be needed in the proof of Proposition \ref{prop:4.1},
 the main ``percolation'' estimate that is the core of our method. The final subsection will complete the proof, thereby establishing exponential decay of the probability that $x$, $y$ lie in the same resonant block. 

\subsubsection{Graphical Sums} \label{4.2.1}

Before getting into a discussion of resonance probabilities, we need to understand more about how
to sum over multi-scale graphs $g_{j'}$.  The goal is to replace any sum of graphs with a
corresponding supremum, multiplied by a factor $c_D^{|g_{j'}|}$.  Graphical sums occur both in the ad expansion for the effective Hamiltonian and in estimates for percolation probabilities. 

If a graph executes
an ordinary step in the lattice, a factor $2D$ will account for the number of choices.  If we have
a jump step from $x$ to $y$, a factor $(2D)^{|x-y|^{(j)}}$ can be used -- this overcounts the number
of possibilities, but matches against the power of $\varepsilon^s$ or $J_0/\varepsilon$ that is
available for bounds on resonance probabilities or perturbative expansion links.  We also pick up
factors of $\exp(ML_{i}^{2/3})$ when the walk passes through a small block ${\overline b}^{(i)}$.
These arise from the block rotation matrices, which lead to sums over states in blocks, 
as well as sums over lattice sites in blocks that may serve as starting points for walks proceeding onward.
In effect, the coordination number of such vertices can
be very large.  But by construction, the minimum graph length for any step into a ${\overline
b}^{(i)}$ is $L_i$.  Of course, $\exp(ML_{i}^{2/3}) \leq c_{D}^{L_{i}}$, but we need to be
cognizant of the fact that in a multi-scale graph, we cannot get away with repeatedly introducing
factors $c_{D}^{|g_{i'}|}$.  But with the $\frac{2}{3}$ exponent, we see that the combinatoric
factor per step from scale $i$ is actually $[\exp(ML_{i}^{-1/3})]^{|g_{i'}|}$.  Then noting that $L_i =
(\tfrac{15}{8})^{i}$, the sum of $L_{i}^{-1/3}$ converges, and the overall combinatoric factor from
blocks of all scales is bounded by $c_{D}^{|g_{j'}|}$.

There are other counting factors that need to be considered.  For example, in each step the
expansion of $(\ad A)^n J$ produces a sum of $n+1$ terms as in (\ref{(2.20a)}).  We also need to sum on
$n$.  There is also the choice of whether to take a jump step or a regular step as we need to
consider both alternatives in (\ref{(3.10)}).  Overall, the number of choices is bounded by $c^n$.
Again, since the minimum graph length for a step on scale $i$ is $L_i$, the overall combinatoric
factor is $\Pi_{i}c^{|g_{j'}|L_{i}^{-1}} \leq \tilde c^{|g_{j'}|}$.  We can see the power of
quadratic convergence (or in our case convergence with exponent $\frac{15}{8}$) in controlling
combinatoric factors.  If $L_i$ grew only linearly with $i$, the combinatoric factors would grow
without bound.

\subsubsection{The Jacobian} \label{4.2.2}

Our method for estimating probabilities of resonant graphs involves a Markov inequality, taking an
expectation of a graph to the $s$ power, and making a bound in terms of a product of
one-dimensional integrals.  In the first step, we could take the integration variables to be
the energy denominators $v_i - v_j$, as long as they form a tree graph.  In the second step,
some denominators involved block energies $E^{(i')}$, but as these moved in sync with uniform
shifts of $v$ in blocks, they could be used as independent integration variables as well.  To
continue this process we have to allow for energy corrections that were moved into
$H_{0}^{(j')}$ from $J^{(j)\ext}$ at the end of step $j \geq 2$.  These terms are bounded like
any other interaction terms as in (\ref{(4.7)}) -- they just happen to be diagonal.  However, if energies
$E^{(i')}$ are to be used as integration variables, we need to control the Jacobian for the
change of variable.  Energy correction terms for $E^{(i)}_{x}$ depend on energies
$E^{(i-1)'}_{y}$ for $y$ up to $L_i /2$ steps away (the graph $g_{(i-1)'}$ has to loop back
to $x$ and contain $y$).  So we need to consider the product of Jacobians $\partial
E_{x}^{(i)}/\partial E_{y}^{(i-1)'}$ over $i \leq j$.  Here $x,y$ are restricted to the particular
graph $g_{j'}$ whose expectation  we are trying to bound.  As explained above in the case of
block energies $E^{(1')}$, all but one of the integration variables in a block ${\overline
b}^{(i)}$ can be replaced with energy differences, and the remaining variable can be shifted to
the energy $E^{(i')}_{x}$ that appears in $g_{i'}$.  The variable $E^{(i')}_{x}$ then
represents constant shifts in energy throughout the block.  In this way, all of the energy
denominators in $g_{i'}$ can be used as integration variables, with the Jacobian for each
intra-block change of variable equal to 1.  Everything is contingent on the denominator graph
$g_{i'}^{\ddd}$ being loop free, with any returns to a block counted as a loop.  We can, however,
work on a loop-free subgraph of $g_{i'}^{\ddd}$.

We now show how to bound $\det {\partial E^{(i')}_{x}}/{\partial E_{y}^{(i-1)'}}$.  
Here $x,y$
belong to the subgraph of $g_{j'}$ corresponding to the level $i$ steps of $g_{j'}$.  (Note that
energy denominators produced in steps $1, \ldots, i-1$ retain their step indices -- see (\ref{(3.3)}) for
a simple example.  
Assume, for the moment, that $x,y$ are site variables, \textit{i.e.} they do not belong to blocks.
Energy correction terms in $E_{x}^{(i')}$ are bounded as in (\ref{(4.7)}).  When
differentiated with respect to $E^{(i-1)'}_{y}$, each term is replaced with a sum of terms with
one of the denominators containing $E^{(i-1)'}_{y}$ being duplicated.  The extra denominator can be bounded as in (\ref{(4.11)}I).  
Thus the derivative can be bounded by $(c
J_0/\varepsilon^{2})^{|g_{(i-1)'}|}$, where the constant $c$ is inserted to account for the sum
over the denominators containing $E_{y}^{(i-1)'}$.  
Summing over $g_{(i-1)'}$ that go from $x$
to $x$ via $y$, we obtain a bound
\begin{equation}
\Delta_{xy} \equiv  \frac{\partial E_{x}^{(i')}}{\partial E_{y}^{(i-1)'}} - \delta_{xy} \leq \sum\limits_{g_{(i-1)'}\, \text{containing}\ x,y} \Big(\frac{c
J_{0}}{\varepsilon^{2}}\Big)^{|g_{(i-1)'}|}, \label{(4.12)}
\end{equation}
which applies as well to $|\Delta_{xy}|$.

If block variables are involved, then there is a complication because the energies are determined through a
two-step process. First, the graphical expansions determine a shift in the effective Hamiltonian of the 
block. Second, the energy shifts are determined by the change in the eigenvalues when the block is 
``rediagonalized.'' But Weyl's inequality implies that the eigenvalues are Lipschitz continuous 
in the matrix entries. The leading term in the map from the 
variables $\{E_{y}^{(i-1)'}\}$ to the variables $\{E_{x}^{(i)'}\}$ is the identity matrix. Hence
the map is bi-Lipschitz, and by Rademacher's theorem the Jacobian is well-defined almost everywhere, 
and the Lipschitz constant bounds the partial derivatives in the Jacobian matrix. The usual 
change of variable formula holds in this context \cite{Evans1991}. Thus 
the argument above applies as well to the cases involving block variables.
Intuitively, one needs control of the measure of regions determined by level surfaces of $\{E_{x}^{(i)'}\}$ for the inverse map; Lipschitz continuity is sufficient for this.

Let us assume that the matrix indices $x,y$ run over a set of $n$ sites/states.  Normally, these
are the vertices of some graph $g_{j'}$, or a subgraph.  From the discussion above
on the combinatorics of graphical sums, we see that row and column sums of $\Delta_{xy}$ are
bounded by $(\tilde c J_0 /\varepsilon^{2})^{L_{i}}$, because $|g_{(i-1)'}| \geq L_i$.  Note that
all $g_{(i-1)'}$ contribute here, not just subgraphs of $g_{j'}$.  We conclude that all
eigenvalues of $I + \Delta$ are in a correspondingly small neighborhood of 1.  Therefore,
\begin{equation}
\left| \log \, \det \bigg[ \frac{\partial E_{x}^{(i')}}{\partial E_{y}^{(i-1)'}}\bigg]\right| \leq n
\bigg(\frac{\tilde c J_0}{\varepsilon^{2}}\bigg)^{L_{i}}.  \label{(4.13)} 
\end{equation}
Due to the rapid growth of $L_i$, this can be summed on $i \geq 2$ to give a bound $e^{J_{0}n}$
on the product of all the Jacobian determinants incurred in using energy denominators as
integration variables.  This bound can easily be absorbed into probability estimates, which are
exponentially small in $n$.     

\subsubsection{Resonant Graphs} \label{4.2.3}

We are now ready to estimate probabilities of resonant graphs as in (\ref{(3.4)}).  First, let us
consider a graph $g_{j'}$ with no returns.  Then the denominator graph likewise has no loops.  (In the
absence of jump steps it has the same number of links as the spatial graph, and the same
vertices.  See below for a general argument.)  Then 
\begin{align}
P \left( A^{(k)\prov}_{xy} (g_{j'}) > (J_0/\varepsilon)^{|g_{j'}|} \right) &\leq \mathbb{E}\left((A^{(k)\prov}_{xy}(g_{j'}))^s  /
(J_0 /\varepsilon)^{s|g_{j'}|} \right)\notag \\
&\leq \varepsilon^{s|g_j'|} \, \mathbb{E} \prod\limits_{uv \in G_{k}^{\ddd}} \big|E_{u}^{(i')} - 
E_{v}^{(i')}\big|^{-s}.   \label{(4.14)}
\end{align}
Here $G_{k}^{\ddd}$ is the denominator graph for $A^{(k)}_{xy}(g_{j'})$, which is $g^{\ddd}_{j'}$ plus
the denominator for $xy$.  Note that the power of $J_0$ in $A_{xy}^{(k)\prov}(g_{j'})$ equals
$|g_{j'}|$, so the $J_0$'s factor out of (\ref{(4.14)}).  The second bound of (\ref{(4.5a)}) ensures that jump
steps contribute their share of $J_0$ factors.  After the change of variable discussed above,
each $E_{u}^{(i')} - E_{v}^{(i')}$ is an independent variable.  We integrate each link, and find  
that (\ref{(4.14)}) is bounded by $(\rho_1
\varepsilon^{s})^{|g_{j'}|}$ (we absorb factors of $e^{J_{0}}$ from the Jacobian 
into $\rho_1$).  The probability for condition I of (\ref{(4.11)}) can likewise be bounded by
$( \rho_1 \varepsilon^s)^{|g_{i'}|}$ by a similar Markov inequality for the single denominator
$E_{x}^{(j')} - E_{y}^{(j')}$ or more simply by noting the length of the integration domain
where (\ref{(4.11)}I) holds.

Now we need to consider the general case for (\ref{(4.11)}II), with $g_{j'}$ ``nearly self-avoiding,''
\textit{i.e.} $|x-y|^{(j)} \geq \frac{7}{8} |g_{j'}|$.  Any return to a site or block will necessarily
shorten the total distance $|x-y|^{(j)}$, because of the ``wasted'' steps.  We can make this
quantitative by drawing a timeline for the walk (graph) $g_{j'}$, with ordinary steps counting
as one time unit, and jump steps of length $m$ counting as $m$ time steps. 
Any time the walk returns to a site/block, we draw an ``arch'' connecting the return time to the
time of the first visit.  The arch graph breaks into connected components, with all sites/blocks
on the graph between the components being visited exactly once.  This is similar to the lace
expansion for self-avoiding walks \cite{Brydges1985}.
Each component represents a time interval during which the walk is executing loops.

\begin{figure}[h]
\centering
\includegraphics[width=.9\textwidth]{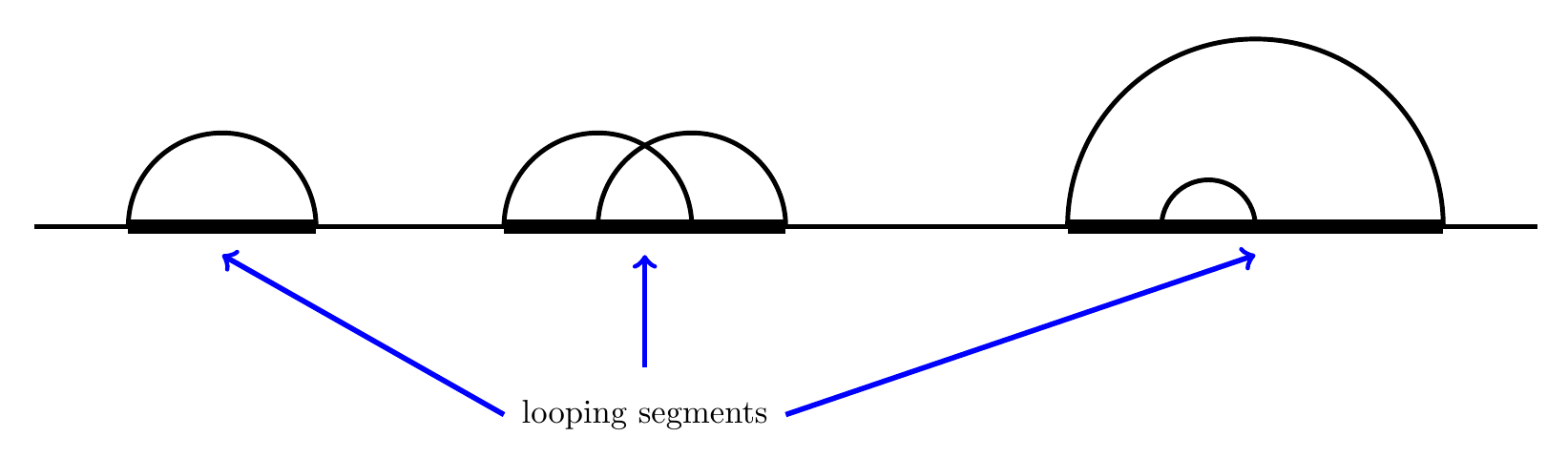}
\caption{Timeline of the walk. Arches connect pairs of times where the walk is at the same
site/block.\label{fig6}}
\end{figure} 

A simple loop/arch of length $\ell$ will cut back the distance traveled by $\ell$.  More
generally, a looping segment of length $\ell$ will cut back the distance traveled by at least
$\frac{2}{3} \ell$.  This is because the spatial graph of the looping portion of the walk is
triply connected.  That is, any surface separating $u$ (the starting point of the looping
section) from $v$ (the final point) will be crossed at least three times by the walk.
(Topologically, the number of crossing must be odd, and a singlet crossing would disconnect the
segment.)  As a result, the length of the graph within the looping segment must be at least
three times the distance from $u$ to $v$, so $\frac{2}{3}$ of the steps are ``wasted''.  We can conclude that the sum of the lengths of the
looping segments cannot be greater than $\frac{3}{2} \cdot \frac{1}{8}
|g_{j'}|=\frac{3}{16}|g_{j'}|$.

\begin{figure}[h]
\centering
\includegraphics[width=.6\textwidth]{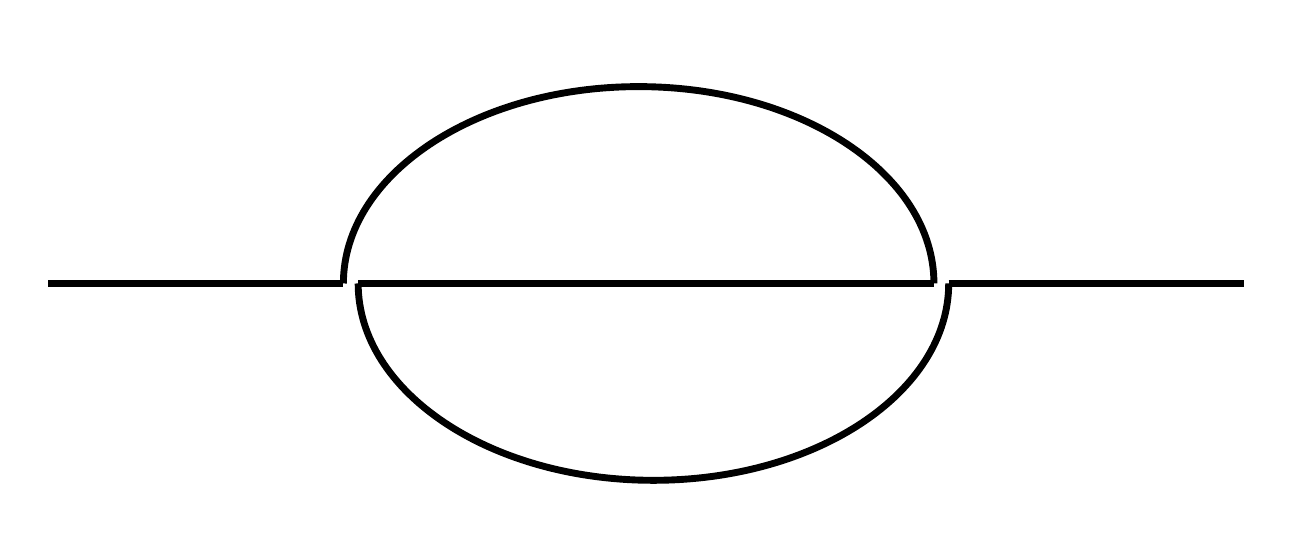}
\caption{A walk executing loops exhibits triple connectivity.\label{fig7}}
\end{figure} 

The next step is to identify certain time intervals containing the looping segments where
inductive bounds (non-probabilistic) will be used in place of Markov inequality bounds.  We will
need to keep a reasonable fraction of the timeline out of the covering intervals, otherwise we
will not get the needed probability decay with $|g_{j'}|$.  The denominator graph has long-range
links, so looping segments will affect the character of the denominator graph in some
neighborhood.

If we consider the denominator graph prior to the identification of vertices on the timeline, it
is devoid of loops.  This is because each time a denominator is produced, 
it connects one or
more loop-free graphs $A$ to a disconnected, loop-free graph $J$.  
(More precisely, for a graph $g_{j'}$ of $J$ that goes from $x$ to $y$, $g_{j'}^{\ddd}$ does not connect $x$ to $y$, so the new denominator cannot create a loop.)
Some denominators are
dropped when they are incorporated into jump steps, but this does not 
spoil the loop-free property.  
(It is useful to keep in mind the ``nested'' character of the denominator links. Graphs 
are constructed as ``walks of walks,'' so the denominator $xy$ in $A^{(i+1)}_{xy}(g_{i'})$ 
encompasses all the previous ones in $g^{\ddd}_{i'}$ on the timeline of $g_{i'}$.)
Let $I_{\alpha}$ be the $\alpha^{\mathrm{th}}$ looping segment of the timeline,
and let $|I_{\alpha}|$ be its length. Let
$\ell=\max_{\alpha} I_\alpha$ and let $i$ be such that $\ell \in[L_{i-1}, L_i)$.  Let
us consider the denominator subgraph $\calD_i$ formed by the links introduced in step $i$
and afterwards, with timeline length in the range $[L_i,L_{i+1})$. As a subgraph of a loop-free graph, $\calD_i$ is of course loop free.  
Furthermore, even after the
identification of sites within looping intervals, it remains loop-free.  This is because each
denominator connects sites at least $L_i$ apart on the timeline, while identifications only
occur within disjoint intervals of length $\leq \ell < L_i$.

\begin{figure}[h]
\centering
\includegraphics[width=.9\textwidth]{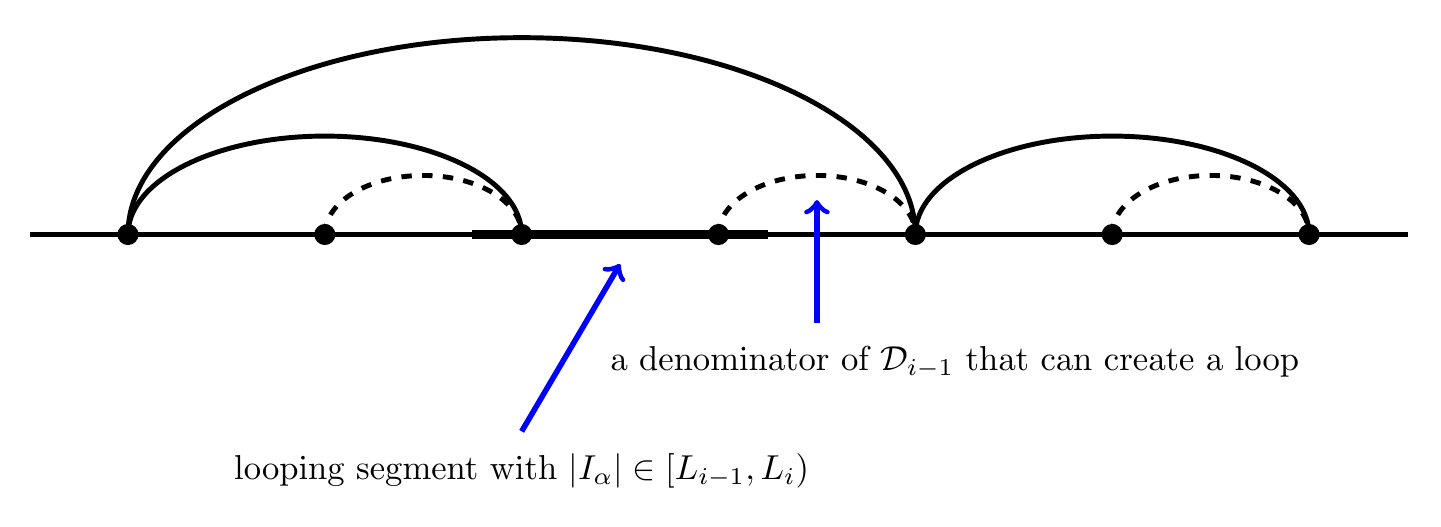}
\caption{The looping segment is too short to spoil the loop-free property of the denominator graph $\calD_i$ (solid arches). When denominators in $\calD_{i-1}$ are added (dashed arches), one is dangerous because it can form a loop after identifying the sites of $I_\alpha$.\label{fig8}}
\end{figure} 

Next, consider what happens when we add denominators from the 
$(i-1)^{\mathrm{st}}$ step, connecting points with timeline separations
in $[L_{i-1},L_i)$. Some of these may be internal to one of the looping segments,
and we will have to replace the corresponding $A$'s by uniform bounds
$(J_0/\varepsilon)^{|g|}$ from (\ref{(4.5a)}). The denominator is effectively erased from the 
denominator graph, along with all denominators nested inside.
A denominator with only one endpoint in a segment $I_\alpha$ is dangerous if
$|I_\alpha| \ge L_{i-1}$ because it could link back indirectly to another point in the
segment through denominators on scales $\ge i-1$.
Therefore, for such a denominator, we replace the corresponding $A$ by its bound
$(J_0/\varepsilon)^{|g|}$. It is not necessary to erase denominators on both sides of 
$I_\alpha$, because $I_\alpha$ can produce at most one identification of sites in the
denominator graph at this stage. 
(The interval $I_\alpha$ cannot contain more than two vertices of $\calD_{i-1}$, because a third would force it to have length $\ge 2L_{i-1} > L_i$.)
 Hence the removal of one denominator link
is sufficient to restore the loop-free property. Through this construction, we obtain
a loop-free denominator graph $\calD_{i-1}$, consisting of non-erased
denominators on scales $\ge i-1$.
We continue the process to smaller length scales $i-2$, $i-3$,..., and
the scale of the ``erased'' denominators never exceeds the scale of the looping
segment it originates from. When the process concludes, we obtain a loop-free
denominator graph $\calD_1$. Each looping segment has a collar of erased
sections of width $\le L_i$ on one side, and $\le L_{i-1}$ on the other, for
$|I_\alpha| \in [L_{i-1},L_i)$. The looping interval ``spoils''
a neighborhood of size no larger than 
$|I_\alpha|+L_i+L_{i-1} \le (2+ \frac{15}{8})|I_\alpha|= \frac{31}{8}|I_\alpha|$.
The total length  of the ``spoiled'' intervals where non-probabilistic bounds are
employed is $\leq \frac{31}{8} \sum\limits_{\alpha} |I_\alpha| \leq \frac{31}{8} \cdot
\frac{3}{16}|g_{j'}| < \frac{3}{4}|g_{j'}|$.

We return to the probability bound as in (\ref{(4.14)}), only now we allow any graph with
$|x-y|^{(j)} \geq \frac{7}{8}|g_{j'}|$.
We obtain
\begin{equation}
P\left(A^{(k)\prov}_{xy} (g_{j'})> (J_0/\varepsilon)^{|g_{j'}|}\right) \leq(\rho_1 \varepsilon^s)^{|g_{j'}|/4}.
\label{(4.15)}
\end{equation}
The ``erased'' sections of $g_{j'}$ contribute factors of $J_0/ \varepsilon$ instead of $J_0$ in
the expectation, so they contribute no smallness to the probability estimate.  But at least
$\frac{1}{4}$ of $g_{j'}$ is clear of looping problems, and so we are able to glean $|g_{j'}|/4$
factors of $\varepsilon^s$ in the Markov inequality.

\subsubsection{Block Probabilities.} \label{4.2.4}

The basic resonance probability bound (\ref{(4.15)}) ensures a positive density of factors of $\varepsilon$
on the walk from $x$ to $y$ used to estimate $P(\calE_{xy}^{(k)})$, the probability that $x$, $y$ belong to the same resonant block $b^{(k)}$ or $B^{(k)}$.  Following the proof of
the analogous bound (\ref{(3.6)}) for $k=2$, we pick up powers of $\varepsilon$ as the walk traverses resonant 
blocks on different scales, and resonant graphs on the current scale.  Each step of a resonant
graph produces a factor $\varepsilon^{s/4}$ from (\ref{(4.15)}), but this becomes $\varepsilon^{s/8}$ because of
the square root in the non-overlapping graph construction.  There is a further degradation of
decay due to the $k^{\mathrm{th}}$ scale collar, which has width $L_{k+1}$.  The minimum length of a
resonant graph $g_{j'}$ is $L_k$, so the density of factors of $\varepsilon^{s/8}$ is reduced by a
factor $L_k/(L_k+2L_{k+1})=\frac{8}{38}$, leaving a residual factor $\varepsilon^{s/38}$ per link.
This would be the uniform rate of probability decay, except for the fact that large
resonant blocks from
earlier scales have their collars increased to $L_{k+1}$.  The increase in the $i^{\mathrm{th}}$ step is
from $L_i$ to $L_{i+1}$, for a total increase along the walk of $2(L_{i+1} - L_i)<2L_i$.  The
diameter of the block is at least $d_i=\exp(2L_i^{2/3})$, so the density of factors of
$\varepsilon^{s/38}$ is reduced slightly by a factor of $d_i / (d_i +2L_i)$ in the $i^{\mathrm{th}}$ step.  The
rapid growth of $d_i$ with $i$ ensures that the density of factors of $\varepsilon$ does not drop
below $\frac{s}{40}=\frac{1}{50}$. (We treated step 2 explicitly in step 2, and subsequent steps have a minor
effect due to the large volumes involved.)  After summing over the graphs in the walk from $x$
to $y$ with the combinatoric bounds established above, we obtain the following result:
\begin{proposition}\label{prop:4.1}
Let $\varepsilon = J_0^{1/20}$ be sufficiently small. Then
\begin{equation}
P(\calE_{xy}^{(k)}) \leq (c_{D}^{50} \rho_1 \varepsilon)^{|x-y|/50}. \label{(4.16)}
\end{equation}
\end{proposition}

\subsection{Perturbation Step and Proof of Inductive Bounds.} \label{4.3}

Let us repeat the analysis of Section \ref{3.2}, writing
\begin{align}
J^{(j')}&=J^{(j')\per} + J^{(j')\res}, \label{(4.17)}
\\
J^{(j')\per}_{xy} &= \sum\limits_{g_{j'}:x \rightarrow y,\, L_{k} \leq |g_{j'}|<L_{k+1},\,
g_{j'} \cap S_{k}=\varnothing,\, B(x) \ne B(y)} J_{xy}^{(j')} (g_{j'}), \label{(4.18)}
\\
A_{xy}^{(k)} (g_{j'})&=\frac{J_{xy}^{(j')\per}}{E_{x}^{(j')}-E_{y}^{(j')}} =
\sum\limits_{g_{j'}} A_{xy}^{(k)}(g_{j'}). \label{(4.19)}
\end{align}
As before, we resum all terms from long graphs with $|g_{j'}|>\frac{8}{7} |x-y|^{(j)}$.  Let
$g_{j''}$ denote either a short graph or a jump step representing all resummed terms.  Then let
\begin{equation}
A_{xy}^{(k)}(g_{j''})=\begin{cases}A_{xy}^{(k)}(g_{j'}), \ \text{if}\ g_{j''}=g_{j'}, \, \text{a
short graph};\\ \sum\limits_{\mathrm{long}\, g_{j'}:x \rightarrow y} A^{(k)}_{xy}(g_{j'}), \
\text{if} \ g_{j''} \ \text{is long.} \end{cases} \label{(4.20)}
\end{equation}
With $\Omega^{(k)}=e^{-A^{(k)}}$, we obtain
\begin{equation}
H^{(k)}=H_0^{(j')} + J^{(j')\res}+J^{(j)\lint} +J^{(k)}. \label{(4.21)}
\end{equation}

We prove our inductive bounds (\ref{(4.5a)}, (\ref{(4.7)}) for $k=j+1$.  For short graphs, we claim that
\begin{equation}
|A^{(k)}_{xy} (g_{j'})|\leq (J_0/\varepsilon)^{|g_{j'}|}. \label{(4.22)}
\end{equation}
To see this, replace jump step subgraphs with their upper bounds from (\ref{(4.5a)}). This transforms $|A^{(k)}_{xy} (g_{j'})|$ into $A^{(k)\prov}_{xy} (g_{j'})$ -- see definition (\ref{(4.10)}). Then $A^{(k)\prov}_{xy} (g_{j'})$ is bounded because $g_{j'}$ is nonresonant, \textit{c.f.} condition (\ref{(4.11)}II). 
For long graphs, we bound numerator and
denominator separately in (\ref{(4.19)}).  The inductive bound (\ref{(4.7)}) applies to the numerator, and the
resonant condition (\ref{(4.11)}I) bounds the denominator from below.  As a result, we have
\begin{equation}
|A_{xy}^{(k)}(g_{j'})|\leq (J_0/\varepsilon^2)^{|g_{j'}|}. \label{(4.23)}
\end{equation}
After summing over long graphs from $x$ to $y$, we obtain
\begin{equation}
|A^{(k)}_{xy}(g_{j''})|\leq(c_{D} J_0/\varepsilon^2)^{\frac{8}{7}|x-y|^{(j)}\lor L_{k}}, \label{(4.24)}
\end{equation}
because all long graphs have $|g_{j'}|\geq \frac{8}{7}|x-y|^{(j)} \lor L_{k}=\frac{8}{7}
|g_{j''}|$, see (\ref{(4.5b)}).
Recall that $\varepsilon=J_{0}^{\delta}$ with $\delta=\frac{1}{20}$.  So $c_{D}^{8/7}
J_{0}^{1/7} \varepsilon^{-16/7} < 1$, and we obtain
\begin{equation}
|A_{xy}^{(k)}(g_{j''})|\leq J_{0}^{|g_{j''}|}, \label{(4.25)}
\end{equation}
which completes the induction for $A^{(k)}$.  Note that this proof and bound applies also to the sum of the absolute values of long graphs. (Since (\ref{(4.25)}) is a stronger estimate, we
have $|A_{xy}^{(k)}(g_{j''})|\leq (J_0/\varepsilon)^{|g_{j''}|}$ for all $g_{j''}$.)  Let us now
examine $J^{(k)}$, which involves terms with a $J^{(j')}$ and one or more $A^{(k)}$ factors.
Combining (\ref{(4.7)}) with the bounds just proven for $A^{(k)}$, we obtain the estimate 
\begin{equation}
|J_{xy}^{(k)}(g_{k})|\leq J_0(J_0/\varepsilon)^{|g_k|-1}, \label{(4.26)}
\end{equation}
which will lead to a proof of (\ref{(4.7)}) for $k = j + 1$ after the block rotations are performed.  Note
that by (\ref{(4.5b)}), the minimum size of $g_{j''}$ in an $A^{(k)}$ term is $\frac{7}{8} L_{k}$.  The
minimum size of a $J^{(j')}$ graph is $L_k$.  Combining these, we obtain a minimum size of
$\frac{15}{8}L_k=L_{k+1}$ for graphs $g_k$. 

\subsection{Diagonalization and Conclusion of Proof} \label{4.4}

As in Section \ref{3.3}, we reorganize terms, putting
\begin{equation}
J^{(k)}+J^{(j')\res}+J^{(j)\lint}=J^{(k)\ext} + J^{(k)\sint}+J^{(k)\lint}. \label{(4.27)}
\end{equation}
Terms whose graph intersects $S_k$ and is contained in ${\overline S}_k$ are put in
$J^{(k)\sint}$ (small block terms) or $J^{(k)\lint}$(large block terms).  Diagonal/intrablock terms for sites/blocks
in $S_{k}^{\textrm{c}}$ are included in $J^{(k)\sint}$ if they are of order less than $L_{k+1}$.  This is
so that in the next step, commutators will not produce terms whose order is less than what is required.

Let $O^{(k)}$ be the matrix that diagonalizes small blocks. Then
\begin{equation}
H_0^{(k')} = O^{(k)\tr} (H_0^{(j')} + J^{(k)\sint})O^{(k)} \label{(4.28)}
\end{equation}
is the new diagonal part of the Hamiltonian.  Then put
\begin{equation}
H^{(k')} = O^{(k)\tr} H^{(k)} O^{(k)}= H_0^{(k')} +J^{(k')} +J^{(k)\lint}.\label{(4.29)}
\end{equation}
Here $J^{(k')}$ is the rotated version of $J^{(k)\ext}$; it has a graphical expansion
with bounds as in (\ref{(4.6)}), (\ref{(4.7)}).

Let us examine the cumulative rotation $R^{(k')} = R^{(j')} \Omega^{(k)} O^{(k)}$ and prove the
``eigenfunction correlator'' estimate, as claimed in (\ref{(4.8)}):
\begin{proposition}
Let $\varepsilon = J_0^{1/20}$ be sufficiently small. Then
\begin{equation}
\mathbb{E} \, \sum\limits_{\alpha} |R_{x\alpha}^{(k')} R_{\alpha y}^{(k')}|\leq
(c_{D}^{50} \rho_1 \varepsilon)^{|x-y|/50}. \label{(4.30)}
\end{equation}
\end{proposition}

\textit{Proof.} We proceed as in the proof of Proposition \ref{prop:3.2}.
The graphical expansions for the matrices $\Omega^{(i)}$, $i \leq k$, lead to walks
that extend from $x$ to $y$ with decay constant $J_0/\varepsilon$.  
With a partition of unity
argument as in (\ref{(2.27)})-(\ref{(2.29)}), the gaps in this walk due to blocks can be filled in with
``probability walks'' along resonant graphs, as in the proof of Proposition \ref{prop:4.1}.    As discussed in the proof of Proposition \ref{prop:3.2}, small block rotations $O^{(j)}$ are combined when not separated by perturbative graphs. The net result is a
combination walk from $x$ to $y$ with a minimum density $\frac{1}{50}$ of factors of
$\varepsilon$. 
All of the walks and state sums are under control, from the discussion on graphical sums in Subsection \ref{4.2.1}. \qed

\textit{Proof of Theorem \ref{thm:1.1}.}
If we let the procedure run to $k=\infty$, off-diagonal matrix elements vanish in the limit.
Then the eigenvalues of the starting Hamiltonian $H^{(\Lambda)}$ are given by the diagonal elements of
$H_0^{\infty}\equiv \lim_{k\rightarrow \infty} H_0^{(k')}$.  They are almost surely nondegenerate, by the argument given in Section \ref{2.3}.   Note that
block formation has to stop eventually in a finite volume $\Lambda$, and after that, 
the effective Hamiltonian $H^{(k')}$ converges rapidly with $k$. Indeed, the off-diagonal entries of $H^{(k)}$ decay exponentially with $L_k = (15/8)^k$. By Weyl's inequality, changes
in the diagonal entries of $H_0^{(k')}$ are correspondingly small. Hence they
converge rapidly to the eigenvalues of $H^{(\Lambda)}$, which are nondegenerate. Once the off-diagonal entries are much smaller than differences in the diagonal entries, the rotation matrices used in our procedure are correspondingly close to the identity. Hence we can define $R^{(\infty)} \equiv \lim_{k \rightarrow \infty} R^{(k')}$, and the eigenfunctions of $H^{(\Lambda)}$ are given by the columns of this limiting rotation.
By bounded convergence, (\ref{(4.30)}) remains true in the limit. This completes the
proof of Theorem 1.1.  \qed

One could improve considerably on the rate of decay proven here, but we have focused on constructing the simplest exposition of the method, rather than optimizing estimates.

\subsection{Labeling of Eigenfunctions and Infinite Volume Limit} \label{4.5}

Each eigenstate has an abstract label $\alpha$. However, working pointwise in the
probability space, one finds that outside of blocks, there is a one-to-one
correspondence between states and sites of the lattice, because each eigenstate has an
expansion exhibiting the predominance of amplitude at a particular site.  In blocks,
there are potentially some choices that need to be made in assigning labels to states.
Labels are assigned when a diagonalization step is performed in a block.  
As explained earlier, the eigenvalues are nondegenerate, with probability 1.
Therefore, except for a set of measure zero, one can label block states in order of
increasing energy. The percolation estimate (\ref{(4.16)}) establishes the diluteness of blocks,
without which the labeling system would lose its significance. From the estimate (\ref{(1.6)}) on the eigenfunction correlator, we see that
eigenfunctions are exponentially small in the distance from the site that they are associated with,
except for a set of exponentially small probability.

As discussed in Subsection \ref{4.2.2}, eigenvalues have been given convergent graphical
expansions, with exponential decay in the size of the graph, \textit{c.f.} (\ref{(4.7)}).  
Eigenfunctions likewise have a local graphical expansions.  Let us use these expansions to demonstrate almost sure convergence of eigenvalues and eigenfunctions as 
$\Lambda$ increases to $\mathbb Z^{D}$.
Let $\Lambda_K=([-K,K]\cap\mathbb{Z})^D$. 
When considering the $K\rightarrow\infty$ limit, it is convenient to use a $K$-independent definition of resonant blocks. In each step of our procedure, a graph will be considered resonant if it is resonant for any value of $K$. Then, in addition to the usual graphical sums for estimating probabilities of resonances, there is a sum over values of $K$ that lead to distinct resonant conditions for a given graph.
The sum over $K$ can be handled in the same manner as the volume factors $\exp(ML_i^{2/3})$ already taken into account in our estimates (\textit{c.f.} Subsection \ref{4.2.1}). 
(We need to sum over the different ways $\Lambda_K$ can intersect the graph and all the sites/blocks that affect it -- even indirectly through graphical expansions of the energies of the graph.) Thus we maintain bounds as in (\ref{(4.15)}) on the probabilities of these generalized resonant graphs.

With this modified procedure, we may work with a fixed configuration of resonant blocks for all values of $K$. For each block $B$, we choose a canonical method of associating sites of $B\cap \Lambda_K$ with the states of that block, for example by using lexicographic order on the sites of $B\cap \Lambda_K$ and matching them to states in order of increasing energy. We may consider, then, the question of convergence of the eigenvalues and eigenfunctions associated with a particular site $x \in \mathbb{Z}^D$.

\begin{theorem}\label{thm:4.3}
Let $J_0$ be sufficiently small, and assume that the probability distribution of the potentials $\{v_x\}_{x\in\mathbb{Z}^D}$ has bounded support. Let $\{E_x^{(K)}, \varphi_x^{(K)}\}_{x \in \Lambda_K}$ denote the eigenvalues and eigenfunctions of $H^{(\Lambda_K)}$, labeled according to the system described above. Then $E_x^{(K)}\rightarrow E_x$ and $\varphi_x^{(K)}\rightarrow \varphi_x$ exponentially almost surely as $K\rightarrow\infty$, with the limits  satisfying $(H-E_x)\varphi_x = 0$. Furthermore,  $\sum_{x \in \Lambda_M} |\varphi_x(y)\varphi_x(z)|$ converges exponentially almost surely as $M\rightarrow \infty$ for each $y$, $z$, and the limit satisfies
\begin{equation}
\mathbb{E} \, \sum\limits_{x\in \mathbb{Z}^D} \big|\varphi_{x}(y)\varphi_{x}(z)\big|\ \leq J_0^{\kappa|y-z|},
\label{(4.31)}
\end{equation}
for some $\kappa >0$ chosen independently of $J_0$.
\end{theorem}

\textit{Proof.} Compare the graphical expansions of the eigenvalues associated with $B(x)$ in two different boxes, $\Lambda_{K_1}$ and $\Lambda_{K_2}$, with $K_1 < K_2$. The difference involves graphs that extend to $\Lambda_{K_1}^{\text{c}}$. (Jump steps need to be rewritten as sums of constituent graphs so as to isolate the ones extending to $\Lambda_{K_1}^{\mathrm{c}}$.) Consider the event $\calE_K(x)$ in which there exists a path from $x$ to $\Lambda_K^{\text{c}}$ with length less than dist$(x,\Lambda_K^{\text{c}})$, in the metric where blocks are contracted to points. By summing over paths and over configurations of blocks along each path, it should be clear that $P(\calE_K(x))$ decays exponentially 
like $J_0^{\kappa\text{dist}(x,\Lambda_K^{\text{c}})}$ for some $\kappa > 0$. By Borel-Cantelli, there is almost surely a $K_0 > 2|x|$ such that $\calE_K(x)$ fails for all $K>K_0$. Bounds on a given graph are governed by the distance it covers between blocks. Hence, as long as $K_1 > K_0$, 
we obtain bounds that decay exponentially in $\text{dist}(x,\Lambda_{K_1}^{\text{c}})$. This means that
differences between effective Hamiltonians of the block $B(x)$ from the change $K_1 \rightarrow K_2$ are exponentially small, once $K_1 > K_0$. By Weyl's inequality, the same is true for the eigenvalues associated with $B(x)$, in particular for $E^{(K)}_x$, the eigenvalue of $H^{(\Lambda_k)}$ that is associated with $x$.

In order to get a similar statement for the corresponding eigenfunction $\varphi^{(K)}_x$, we need some quantitative control on the gaps between the eigenvalues associated with $B(x)$. For simplicity, we have assumed that the probability distribution of the potentials $\{v_x\}_{x\in\mathbb{Z}^D}$ is supported on a bounded interval. Consider the spectrum of $H^{(\Lambda_K)}$, which is then also supported on a bounded interval. Let $\calF_K$ be the event that there is a gap smaller than $K^{-q}$. From Minami's estimate \cite{Minami1996}, one can show that $P(\calF_K)$  is bounded by a constant times $\rho_0 K^{-q+2D}$, for $q>2D$ -- see \cite{Klein2006}, eq. 8. Taking $q=2D+2$, the probabilities are summable, so by Borel-Cantelli there is almost surely a $\tilde{K_0}>K_0$ such that $\calF_K$ fails for all $K>\tilde{K_0}$. As explained above, differences between corresponding effective Hamiltonians of the block $B(x)$ from the change $K_1 \rightarrow K_2$ are exponentially small, so for $K_1 > \tilde{K}_0$ they are much smaller than the gaps between eigenvalues associated with $B(x)$. (Here we use the fact that these eigenvalues agree with those of $H^{(\Lambda_{K_1})}$ within an exponentially small error, as explained in the proof of Theorem \ref{thm:1.1}.) This implies that the mixing of $\varphi^{(K_1)}_x$ with the other eigenfunctions associated with $B(x)$ are similarly small. Thus we obtain almost sure exponential convergence of both $E^{(K)}_x$ and $\varphi^{(K)}_x$ as $K \rightarrow \infty$.
Observe that $(H-E_x^{(K)})\varphi_x^{(K)} \rightarrow 0$ as $K\rightarrow \infty$ because $H-H^{(\Lambda_K)}$ only affects $\varphi_x^{(K)}$ at the boundary of $\Lambda_K$, and the relevant graphs are exponentially small in $\text{dist}(x,\Lambda_K^{\text{c}})$. Hence $(H-E_x)\varphi_x = 0$ almost surely.

\sloppy
The same arguments can be used to demonstrate almost sure exponential convergence of
$\sum_{x \in \Lambda_M} |\varphi_x(y)\varphi_x(z)|$ as $M\rightarrow \infty$ for each $y$, $z$, because the graphs involved extend from $\{y,z\}$ to $\Lambda_M^{\text{c}}$. By bounded convergence, the eigenfunction correlator estimate of Theorem \ref{thm:1.1} extends to the limits $K$, $M \rightarrow \infty$, completing the proof. \qed

\section*{Acknowledgement}
The author would like to thank Tom Spencer for a collaboration over several years, during which time many of these ideas were developed. Comments from Wojciech De Roeck were  helpful in improving an earlier draft of this work.
\begin{footnotesize}
\bibliographystyle{acm}
%\bibliography{library}

\end{footnotesize}
\end{document}